\documentclass[]{aa}
\usepackage{graphicx}
\graphicspath{{./figures/}}
\usepackage{natbib}
\usepackage{amssymb,amsmath,amsfonts}
\usepackage{epsfig}
\usepackage{mathptmx}
\usepackage{longtable}
\usepackage{comment}
\usepackage{color}
\usepackage[caption=false]{subfig}
\usepackage{animate}
\usepackage{rotating}
\usepackage{pdflscape}
\usepackage{url}
\usepackage{enumitem}
\usepackage[draft]{hyperref}
\usepackage[normalem]{ulem}
\usepackage[utf8]{inputenc}
\usepackage{longtable} 
\usepackage{pdflscape} 
\newcounter{dummy}  
\usepackage{booktabs}
\usepackage{siunitx}
\usepackage{longtable}
\usepackage{threeparttablex}
\usepackage{booktabs}
\usepackage{array}
\newcolumntype{d}[1]{>{\raggedleft\arraybackslash}p{#1em}}

\begin{document} 

\title{The LOFAR Two-metre Sky Survey}
\subtitle{VII. Third Data Release}
\authorrunning{Shimwell et~al.}
\titlerunning{LoTSS-DR3}
\author{}
\institute{}
\date{Accepted January 13, 2026; received October 18, 2025; in original form \today}

\author{T. W. Shimwell$^{\ref{ASTRON}, \ref{Leiden}}$\thanks{E-mail: shimwell@astron.nl},
M. J. Hardcastle$^{\ref{Hertfordshire}}$, 
C. Tasse$^{\ref{LUX},\ref{Rhodes}, \ref{CNRS}}$,
A. Drabent$^{\ref{Tautenburg}}$, 
A. Botteon$^{\ref{INAF}}$, 
W.L.~Williams$^{\ref{SKA}}$,
P.N.~Best$^{\ref{Edinburgh}}$,
H.J.A.~R\"{o}ttgering$^{\ref{Leiden}}$,
M.~Br\"uggen$^{\ref{Hamburg}}$,
G.~Brunetti$^{\ref{INAF}}$, 
J.R.~Callingham$^{\ref{ASTRON},\ref{UvA}}$,
K. T. Chy\.zy$^{\ref{Krakow}}$, 
J.E.~Conway$^{\ref{Chalmers}}$, 
F.~De~Gasperin$^{\ref{INAF}}$, 
M.~Haverkorn$^{\ref{Nijmegen}}$, 
C.~Horellou$^{\ref{Chalmers}}$, 
N.~Jackson$^{\ref{Manchester}}$,
G.K.~Miley$^{\ref{Leiden}}$, 
L.K.~Morabito$^{\ref{Astronomy_Durham}, \ref{Cosmology_Durham}}$,
R.~Morganti$^{\ref{ASTRON}, \ref{Groningen}}$,
S.P.~O'Sullivan$^{\ref{Madrid}}$,
D.J.~Schwarz$^{\ref{Bielefeld}}$,
D.J.B.~Smith$^{\ref{Hertfordshire}}$,
R.J.~van~Weeren$^{\ref{Leiden}}$, 
H.K.~Vedantham$^{\ref{ASTRON}, \ref{Groningen}}$,
G.J.~White$^{\ref{Open_University},\ref{RAL}}$,
A.~Ahmadi$^{\ref{ASTRON}}$,
L.~Alegre$^{\ref{Hertfordshire}}$, 
M.~Arias$^{\ref{ASTRON}}$,
B.~Asabere$^{\ref{ASTRON}}$,
B.~Bahr-Kalus$^{\ref{INAF_Torino},\ref{Torino}, \ref{INFN_Torino}}$,
B.~Barkus$^{\ref{Hertfordshire}}$,
M.~Bilicki$^{\ref{Warsaw_Theoretical}}$,
L.~B{\"o}hme$^{\ref{Bielefeld}}$,
M.~Brentjens$^{\ref{ASTRON}}$, 
M.~Brienza$^{\ref{INAF}}$, 
D.J.~Bomans$^{\ref{Bochum}}$, 
A.~Bonafede$^{\ref{Bologna}, \ref{INAF}}$, 
M.~Bonato$^{\ref{INAF}}$, 
E.~Bonnassieux$^{\ref{Bordeaux}}$,
J.M.~Boxelaar$^{\ref{INAF}, \ref{Bologna}}$, 
S.~Camera$^{\ref{Torino},\ref{INFN_Torino}, \ref{INAF_Torino}}$,
R.~Cassano$^{\ref{INAF}}$, 
J.~Chilufya$^{\ref{Hertfordshire}}$,
M.~Cianfaglione$^{\ref{Bologna}, \ref{INAF}}$, 
J.H.~Croston$^{\ref{Open_University}}$, 
V.~Cuciti$^{\ref{Bologna}, \ref{INAF}}$, 
P.~Dabhade$^{\ref{NCNR_Warsaw}}$,
E.~De~Rubeis$^{\ref{Bologna}, \ref{INAF}}$, 
J.M.G.H.J.~de~Jong$^{\ref{Leiden},\ref{ASTRON}}$,
D.~Dallacasa$^{\ref{Bologna}, \ref{INAF}}$,
R.J.~Dettmar$^{\ref{Bochum}}$, 
K.J.~Duncan$^{\ref{Edinburgh}}$, 
G.~Di~Gennaro$^{\ref{INAF}}$, 
H.W.~Edler$^{\ref{ASTRON}}$,
C.~Groeneveld$^{\ref{INAF}}$, 
G.~G\"urkan$^{\ref{Hertfordshire},\ref{CSIRO}}$,
M.~Hajduk$^{\ref{Olsztyn}}$, 
C.L.~Hale$^{\ref{Edinburgh},\ref{Oxford}}$,
V.~Heesen$^{\ref{Hamburg}}$,
D.N.~Hoang$^{\ref{Tautenburg}}$,
M.~Hoeft$^{\ref{Tautenburg}}$,
H.~Holties$^{\ref{ASTRON}}$,
M.A.~Horton$^{\ref{Cambridge}}$,
M.~Iacobelli$^{\ref{ASTRON}}$, 
M.~Jamrozy$^{\ref{Krakow}}$, 
M.J.~Jarvis$^{\ref{Oxford},\ref{Western_Cape}}$,
V.~Jelic$^{\ref{Zagreb}}$, 
M.~Kadler$^{\ref{Wurzburg}}$,
R.~Kondapally$^{\ref{Astronomy_Durham}, \ref{Cosmology_Durham}}$,
M.~Kunert-Bajraszewska$^{\ref{Torun}}$, 
M.~Loose$^{\ref{ASTRON}}$,
M.~Magliocchetti$^{\ref{INAF-IAPS}}$, 
K.~Ma{\l}ek$^{\ref{NCNR_Warsaw}}$,
C.~Manzano$^{\ref{Juelich}}$,
J.P.~McKean$^{\ref{Groningen}, \ref{SARAO}}$,
M.~Mevius$^{\ref{ASTRON}, \ref{Groningen}}$,
B.~Mingo$^{\ref{Hertfordshire}}$,
A.~Miskolczi$^{\ref{Juelich}}$,
A.~Misra$^{\ref{Krakow}}$,
J.~Mold\'on$^{\ref{Granada}}$, 
D.G.~Nair$^{\ref{Concepcion}, \ref{MPI_Bonn}}$,
S.J.~Nakoneczny$^{\ref{Warsaw_Theoretical}}$,
E.~Orru$^{\ref{ASTRON}}$,
M.~Pashapour-Ahmadabadi$^{\ref{Bielefeld}}$,
T.~Pasini$^{\ref{INAF}}$, 
J.~Petley$^{\ref{Leiden}}$, 
J.C.S.~Pierce$^{\ref{Hertfordshire}}$,
I.~Prandoni$^{\ref{INAF}}$, 
D.~Rafferty$^{\ref{Hamburg}}$,
K.~Rajpurohit$^{\ref{Harvard}}$,
C.J.~Riseley$^{\ref{Bochum}, \ref{Bochum_RAPP}}$,
I.D.~Roberts$^{\ref{Waterloo_Astrophysics},\ref{Waterloo_Physics}}$,
S.~Sethi$^{\ref{Krakow}}$, 
A.~Shulevski$^{\ref{ASTRON},\ref{Groningen}, \ref{UvA}, \ref{Skopje}}$,
M.~Stein$^{\ref{Bochum}}$, 
C.~Stuardi$^{\ref{INAF}}$, 
F.~Sweijen$^{\ref{Astronomy_Durham}}$,
S.~ter~Veen$^{\ref{ASTRON}}$,
R.~Timmerman$^{\ref{Astronomy_Durham}, \ref{Cosmology_Durham}}$,
M.~Vaccari$^{\ref{Inter_Cape_Town},\ref{Inter_Western_Cape}, \ref{INAF}}$ and 
S.~Wijnholds$^{\ref{ASTRON}}$.\\ \vspace{0.5cm}
\textit{(Affiliations can be found after the references)}}

\institute{  }

\abstract{
\noindent
We present the third data release of the LOFAR Two-metre Sky Survey (LoTSS-DR3). The survey images cover 88\% of the northern sky and were created from 12,950\,hrs of data (18.6\,PB) accumulated over 10.5 years. Producing the images took 20 million core hours of processing through direction-independent and direction-dependent calibration pipelines that correct for instrumental effects as well as spatially and temporally varying ionospheric distortions. In our 120-168\,MHz continuum mosaic images with an angular resolution of 6$\arcsec$ (9$\arcsec$ below declination 10$^\circ$) we catalogue 13,667,877 sources, formed from 16,943,656 Gaussian components. The scatter in the astrometric precision approximately follows the expected noise-like behaviour but with an additional systematic component of at least 0.24$\arcsec$ that is likely due to calibration imperfections. The random flux density scale error is 6\%, while the systematic offset  was previously shown to be within 2\%. The median sensitivity of our mosaics is 92$\mu$Jy beam$^{-1}$, improving to 68$\mu$Jy beam$^{-1}$ at high observing elevations, but degrading to 183$\mu$Jy beam$^{-1}$ at the celestial equator due to station area projection effects.  Completeness simulations, accounting for realistic source models, time- and bandwidth-smearing effects, and astrometric errors, indicate that we detect more than 95\% of compact sources with integrated flux densities exceeding 9 times the local root mean square (RMS) noise. However,  the recovered source counts in a particular integrated flux density bin do not match the injected counts until flux densities exceed 45 times the local RMS noise. The Euclidean-normalised differential source counts derived from the survey constrain the radio source population over five orders of magnitude and are in good agreement with previous deep and wide-area surveys. All data products are publicly available, including catalogues, individual-field Stokes I, Q, U, and V images, mosaicked Stokes I images, and $uv$ data with associated direction-dependent calibration solutions.
}

\keywords{surveys -- catalogues -- radio continuum: general -- techniques: interferometric}
 \maketitle

\section{Introduction}

The LOw Frequency ARray (LOFAR; \citealt{vanHaarlem_2013}) Two-metre Sky Survey (LoTSS; \citealt{Shimwell_2017}) is a 120-168\,MHz imaging survey of the entire northern sky. The primary observational aim of the survey is to produce total intensity images with a root mean square (RMS) noise of $\sim$100$\mu$Jy beam$^{-1}$ at an angular resolution of 6$\arcsec$ under optimal observing elevations. However, other science-ready data products are also being produced, including: lower-angular-resolution images with a higher surface brightness (e.g. \citealt{Oei_2022}); Stokes Q and U image cubes at several resolutions (e.g. \citealt{Erceg_2022} and \citealt{OSullivan_2023}); Stokes V images (e.g. \citealt{Callingham_2023}); and dynamic spectra of selected variable sources (e.g. \citealt{Tasse_submitted}). In addition to these products, we provide radio source catalogues and $uv$ data together with the associated calibration solutions. The $uv$ data can be easily post-processed and tailored to a particular science case by, for example, tuning calibration and imaging parameters to study faint diffuse emission (\citealt{vanWeeren_2021}) or creating source-subtracted data and snapshot images to search for transients (\citealt{deRuiter_2024}). Efforts are also ongoing to enhance the radio catalogues with optical identifications, photometric redshifts, and host-galaxy properties for our catalogued radio sources where possible  (see \citealt{Duncan_2019}, \citealt{Williams_2019}, \citealt{Kondapally_2021}, \citealt{Hardcastle_2023}).

Other complementary very wide area radio surveys are either being conducted or have recently been completed. These include the Apertif imaging survey (\citealt{Adams_2022}), the Evolutionary Map of the Universe  (EMU; \citealt{Norris_2011} and \citealt{Hopkins_2025}), the GaLactic and Extragalactic All-sky Murchison Widefield Array survey (GLEAM; \citealt{HurleyWalker_2017}) and its eXtended component (GLEAM-X; \citealt{HurleyWalker_2022}), the LOFAR Decameter Sky
Survey (LoDeSS; \citealt{Groeneveld_2024}), the LOFAR Low-band Sky Survey (LoLSS; \citealt{deGasperin_2021}),  
the Rapid ASKAP Continuum Survey (RACS; \citealt{McConnell_2020}), the TIFR GMRT Sky Survey Alternative Data Release (TGSS-ADR1; \citealt{Intema_2017}), and the Karl G. Jansky Very Large Array Sky Survey (VLASS; \citealt{Lacy_2020}). Between these surveys, sensitive data and detailed images of the entire sky are publicly available across a range of frequencies. Furthermore, substantial sky areas (tens of square degrees) are being probed at an exceptionally high sensitivity and/or angular resolution by the MeerKAT International GigaHertz Tiered Extragalactic Exploration (MIGHTEE; \citealt{Jarvis_2016}) survey and LOFAR Two-metre Sky Survey Deep Fields (LoTSS Deep; \citealt{Best_2023}).  Together, these projects are dramatically increasing the number of known radio sources and enriching our view of the Universe. They are facilitating a vast number of studies covering a wide range of science topics, including: active galactic nuclei; star formation; the intracluster medium; cosmic magnetism; cosmology; transients; pulsars; supernovae; and stars and exoplanets.

\begin{figure}[htbp]
   \centering
   \includegraphics[width=1.0\linewidth]{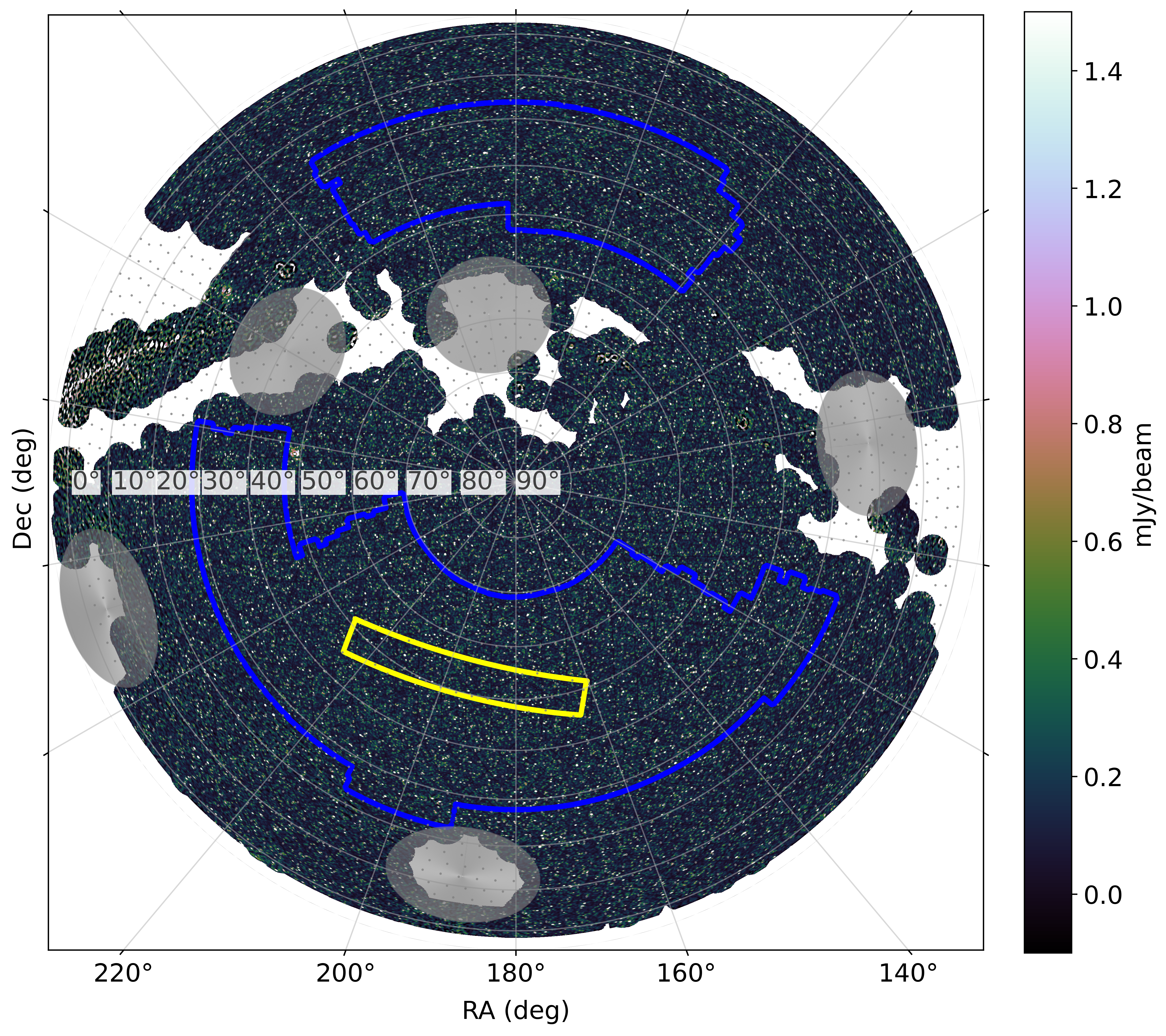}
   \includegraphics[width=1.0\linewidth]{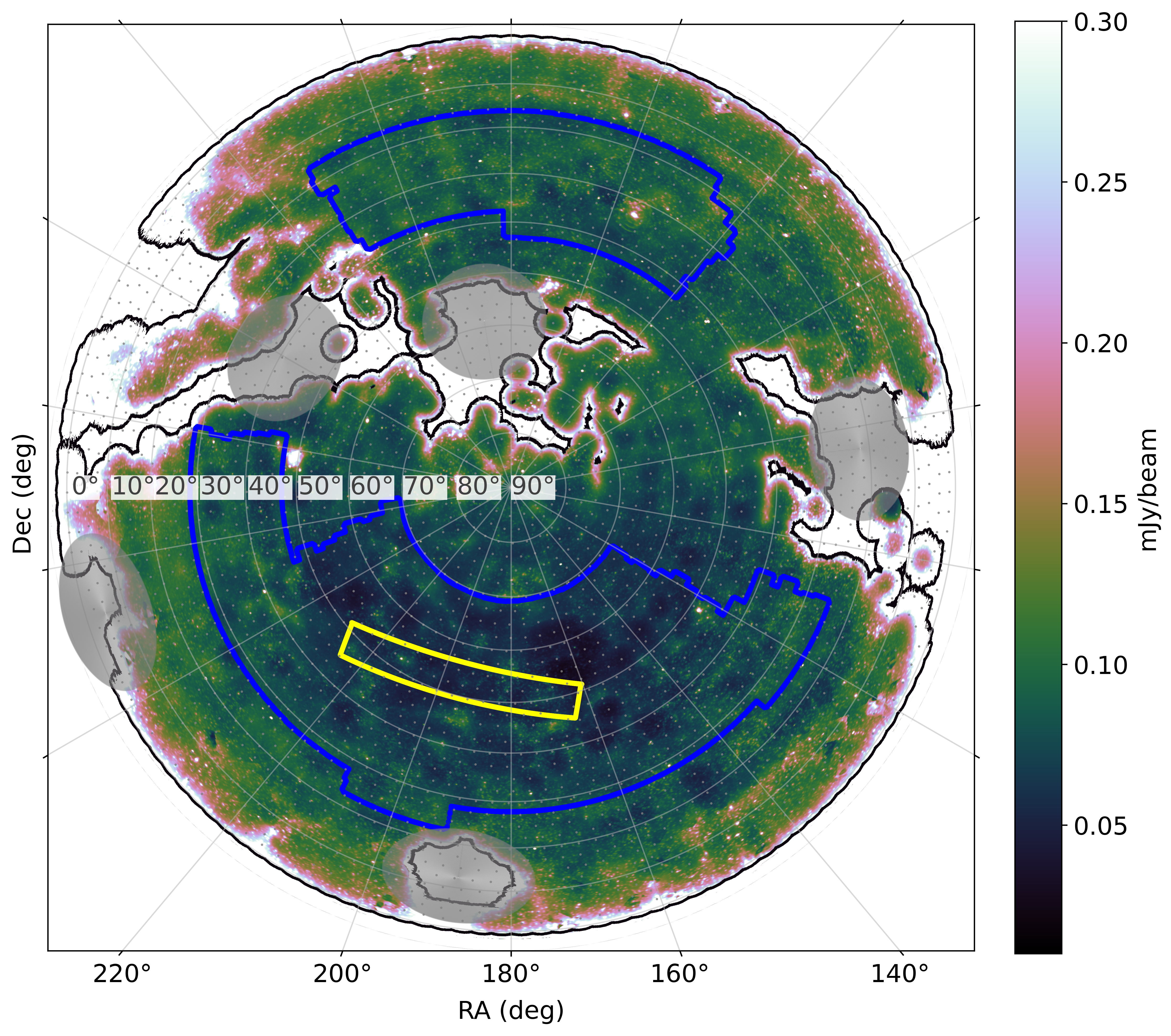}
   \vspace{-0.2cm}
   \caption{Top: Re-projection of the LoTSS-DR3 mosaic images. Bottom: Corresponding RMS image. The yellow and blue outlines show the LoTSS-DR1 and LoTSS-DR2 areas, which cover 2\% and 27\% of the northern sky, respectively. The black outline in the bottom panel shows the LoTSS-DR3 coverage of 88\% of the northern sky. The large grey circles show regions that are within 10$^\circ$ of the bright radio sources Cassiopeia A, Cygnus A, Taurus A, Hercules A, or Virgo A. The small grey dots show the locations of the 3,168 LoTSS pointings of which 2,551 are included in this data release. }
   \label{Fig:allsky_projection}
\end{figure}

Previously, we have made three public LoTSS data releases: the preliminary data release (PDR; \citealt{Shimwell_2017}) and two full-quality data releases (LoTSS-DR1 and LoTSS-DR2; \citealt{Shimwell_2019,Shimwell_2022}) that
provided I, Q, U, and V images, catalogues, and calibrated $uv$ data. Our largest previous data release, DR2, covered 27\% of the
northern sky and contained  4,396,228 radio sources. Here we present the third LoTSS data release (LoTSS-DR3; see Fig. \ref{Fig:allsky_projection}), which spans 88\% of the northern sky and catalogues 13,667,877 radio sources. This
release includes 95\% of the data that we have gathered for the project to date. In Sec. \ref{Sec:observations} we describe the observations and data processing. In Sec. \ref{Sec:individual_pointings} and Sec. \ref{Sec:mosaic_quality} we analyse the individual pointing image quality and mosaic image quality,  respectively. In Sec. \ref{Sec:image_analysis} we derive the Euclidean-normalised differential source counts. We outline some future prospects in Sec. \ref{Sec:future_prospects} and provide a summary in Sec.\ref{Sec:Summary}. Finally, in Sec. \ref{Sec:release_details} we describe the publicly released data products. The spectral index, $\alpha$, is defined throughout the paper as $S_\nu \propto \nu^\alpha$, where $S_\nu$ is the flux density at frequency $\nu$.

\section{Observations and data processing}
\label{Sec:observations}

The existing observations for LoTSS were taken between May 23, 2014, and August 19, 2024. During this period more than 13,575\,hrs of data were recorded for the project, making it one of the largest individual observing projects ever carried out. While there is some variation in the observation properties, LoTSS observations are generally conducted using the full international LOFAR array of 48 core stations (30.8\,m diameter, baselines 40\,m to 3,700\,m), 14 remote stations (only 30.8\,m of the 41\,m diameter used in High Band Antenna, HBA, Dual Inner mode, baselines 2,300\,m to 120\,km) and up to 14 international stations (56.5\,m diameter, baselines 53\,km to 1980\,km). The observations use the available 96\,MHz bandwidth, split equally between two pointing centres, giving each pointing 120-168\,MHz coverage with a central frequency of 144\,MHz. The data are flagged using AOFlagger (\citealt{Offringa_2012}), averaged in DP3 (\citealt{vanDiepen_2018}) and compressed using Dysco with a bit rate of 10 (\citealt{Offringa_2016}) before being archived in the LOFAR Long Term Archive (LTA\footnote{\url{https://lta.lofar.eu/}}) with a time resolution of 1\,s and a frequency resolution of 12.1875\,kHz. Pointings above declination 10$^\circ$ are typically observed for a total of 8\,hrs, whilst pointings below this declination are observed for a total of 12\,hrs. To reduce station area projection effects and maintain good sensitivity, all pointings are observed at high elevation, with a minimum elevation limit of 30$^\circ$. With this limit, normally observations covering the entire required integration time can be conducted for fields above declination 20$^\circ$, whereas fields at lower declinations are generally observed multiple times with shorter integrations (as little as 2\,hrs).

The main variations in the LoTSS observations are due to international station coverage. Only eight international stations were available when observations began in 2014, and for the first year of observations the international stations were not always included in the observations, but after this, they were routinely included. For data without the international stations, the time and frequency averaging was coarser (typically 2\,s and 24.375\,kHz) as these observations did not have to mitigate against the severe time- and frequency-averaging smearing effects that affect the international baselines. The other main variation is integration time: whilst 86\% of fields have within 10\% of the desired integration time, 12\% have additional time (up to a total integration time of 49\,hrs) and 2\% have too little (as low as 4\,hrs). Here the additional observing time on a pointing is mostly due to contributions from co-observing programs,  in which LoTSS pointings are observed simultaneously with another target field (either a LoTSS pointing or otherwise).

When complete, LoTSS will consist of 3,168 pointings, but only 2,665 have been observed so far. Unobserved pointings primarily lie within an absolute Galactic latitude $|b| < 20^\circ$, with fewer missing directly on the Galactic plane. Furthermore, 229 pointings (of which 134 have data) are located within 10$^\circ$ of exceptionally bright radio sources, namely Cassiopeia A, Cygnus A, Virgo A, Taurus A, and Hercules A, which have integrated flux densities between 475\,Jy and 10,700\,Jy at these frequencies (\citealt{deGasperin_2020,Timmerman_2022}) and therefore significantly contaminate the surrounding region (see e.g. \citealt{Edler_2023} who successfully mapped the area around Virgo A using identical data to LoTSS but with an adjusted data processing strategy). The LoTSS-DR3 fields include all regions where data were accumulated before LOFAR observations were suspended in September 2024 for the LOFAR2.0 upgrade\footnote{\url{https://www.lofar.eu/}}, excluding areas within $10^\circ$ of the bright radio sources mentioned above (just 20 observed fields are included in these regions). This leaves 2,551 fields in DR3.

The 12,950\,hrs of data utilised for LoTSS-DR3 were accumulated through
2,720 observing sessions (generally observing two pointings
simultaneously) and 100 different observing proposals. The large
number of contributing proposals comes from co-observing programs that provided data. Such
observations effectively contributed 1,460\,hrs of data towards
LoTSS-DR3 after accounting for the normal LoTSS observing strategy of taking data for two pointings
simultaneously. However, the bulk of the data, 11,490\,hrs, was accumulated through the dedicated LoTSS proposals LC2\_038, LC3\_008, LC4\_034, LT5\_007, LC6\_015, LC7\_024, LC8\_022, LC9\_030, LT10\_010, LC11\_020, LT14\_004, LT16\_012, and LC20\_026. In total, the data that were processed amounts to  18.6\,PB (split over 2.4 million files)  and are stored in the LTA (36\% at SURF in Amsterdam, 56\% at Forschungszentrum J\"{u}lich, and 8\% at PCSS, the Pozna\'n Supercomputing and Networking Center) from where they were retrieved for processing.

The data processing follows the same procedure as for LoTSS-DR2 and the
LoTSS-Deep fields (see \citealt{Shimwell_2022} and
\citealt{Tasse_2021}), and as such it utilises the LOFAR Initial Calibration pipeline (LINC\footnote{\url{https://linc.readthedocs.io/en/latest/}}; \citealt{deGasperin_2019}, previously called PreFactor) for direction-independent calibration, followed
by the DDF-pipeline\footnote{\url{https://github.com/mhardcastle/ddf-pipeline}: versions 2.5, 3.0 and 3.1 were used.} for direction-dependent calibration and imaging. These pipelines make use of various software packages, with LINC using DP3 for calibration and WSClean (\citealt{{Offringa_2014}}) for imaging, whilst DDF-pipeline makes use of kMS (\citealt{Tasse_2014} and \citealt{Smirnov_2015}) for simultaneously deriving calibration solutions in a number of directions and DDFacet  (\citealt{Tasse_2018}) for imaging with the derived solutions applied. LINC first removes the international stations from the data and then uses the calibrator observation to derive direction-independent and time-independent calibration parameters, namely: bandpass, polarisation alignment and clock offsets. It then applies these solutions to the target field data, which is further corrected for the direction-independent ionospheric Faraday rotation measure before the phase is calibrated using the TGSS-ADR1 model of the field or, in some cases, where the TGSS-ADR1 model is inadequate, self-calibrated from a direction-independent calibrated image made from the data. The target data are also flagged for interference, and bright off-axis sources (Cassiopeia A, Cygnus A) are subtracted from the data using the demixing technique (\citealt{vandertol_2007}) if they are within 30$^\circ$ of the target field. In this process, the visibilities are first phase-rotated to the direction of the bright off-axis source and then averaged to decorrelate the target field. The averaged data are calibrated against a model of the bright source, and the resulting gain solutions are used to predict, and subsequently subtract, its contribution to the input visibilities. When the sources are farther from the target field, baselines and time intervals are instead flagged (`A-team clipping') if the predicted signal from these sources (and also Taurus A, Hercules A or Virgo A) exceeds the observatory recommended threshold of 5\,Jy.

In DDF-pipeline, the data are imaged and are corrected
for direction-dependent effects such as beam model imperfections
or ionospheric distortions that are severe at these low radio
frequencies. The imaging accounts for the baseline-, time- and frequency-dependent LOFAR primary beam, and the final images are corrected using the baseline-, time- and frequency-averaged beam response. DDF-pipeline performs a direction-dependent
self-calibration, generally calibrating the data in 45 different directions
(facets) and imaging the data 5 times, and calibrating 4
times, in the process building up a high-quality frequency-dependent (simple power-law) sky model for the field. For fields below declination $10^\circ$ DDF-pipeline first removes sources outside the $8^\circ \times 8^\circ$ region that is imaged to reduce the artefacts caused by them.

In many cases, the LoTSS-DR2 images are directly
reused in LoTSS-DR3 (719 out of 841 pointings) but in other cases, the
images were recreated with refined data processing or to include more
data, where refined data processing involves some combination of better removal of bright off-axis sources (using the demixing technique instead of flagging contaminated data), self-calibration in LINC, or tuning parameters in DDF-pipeline. The vast majority (94\%) of fields were processed with no modifications to the standard DDF-pipeline strategy (phase and amplitude calibration solutions applied in 45 calibration directions) but the remaining 6\% required some refinement. This was generally near bright sources or in regions with bright Galactic emission, where we were able to improve the image quality by minor changes in the DDF-pipeline parameter set such as reducing the number of calibration directions, turning off amplitude corrections, creating refined masks for deconvolution or subtracting sources outside the field of view. 

The LINC processing times varied significantly between observations (from 150  to 4,300 core hours) depending on the observation duration (2-8\,hrs) and whether the data were demixed and/or self-calibrated. In total, we estimate that approximately 4.2 million core hours were used for LINC processing, and this processing was done over a period of about 7 years (early 2018 to early 2025). To reduce data transfer,  LINC processing for data stored at the Forschungszentrum J\"{u}lich LTA site was primarily performed on the local JUWELS cluster, while data stored at the SURF LTA site were processed on the local Spider cluster. Data stored in Pozna\'{n} were processed on both Spider and JUWELS.

\begin{figure}[htbp]
   \centering
   \includegraphics[width=\linewidth]{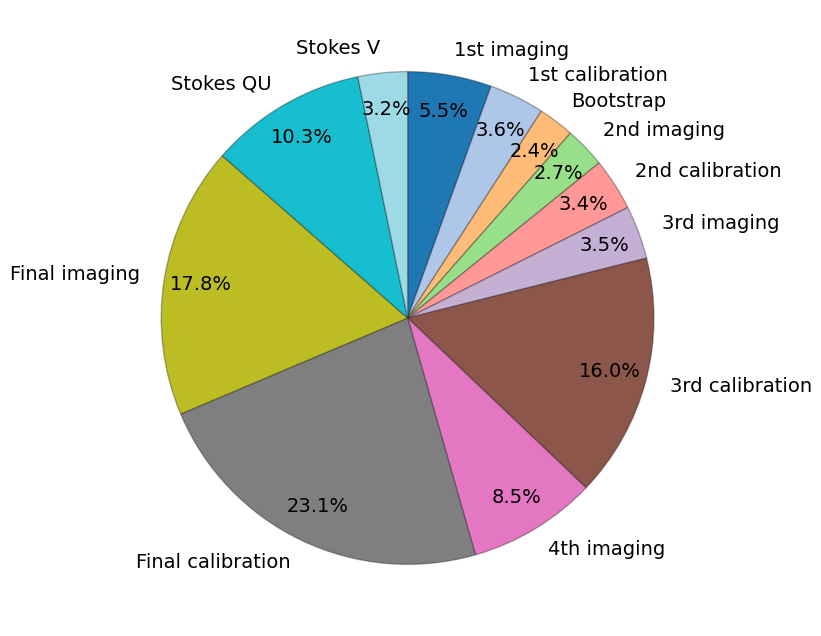}
    \vspace{-0.2cm}
   \caption{Breakdown of the computational time for a typical DDF-pipeline run on a standard 8\,hr observation with 48\,MHz bandwidth. The total runtime is 75\,hrs using 60 CPU cores on a node with AMD EPYC Bergamo processors (128 cores per socket). The runtime includes the creation of all Stokes I, Q, U, and  V science products, five Stokes I imaging cycles, and four calibration cycles.}
   \label{Fig:spider_compute}
\end{figure}

The main computational expense is in the direction-dependent calibration. From the DDF-pipeline log files, we estimate that the data processing took 16 million core hours (an average of 6,200 core hours per field) over approximately the same 7 year period. These estimates are sums taken from all data included in the release, but the processing was performed on a variety of different compute nodes with differing performance. For example, if we look at just the 259 fields processed using 60 CPU cores on a node with AMD EPYC Bergamo processors (128 cores per socket), we typically find that an 8\,hr field with default settings took a median of 75\,hrs (standard deviation of 12\,hrs) to produce a full set of Stokes I, Q, U, and V products, with a breakdown in processing that is shown in Fig. \ref{Fig:spider_compute}. Processing the 145 12-hr fields on the same system we find that they typically took 135\,hrs. DDF-pipeline was run in a fully distributed way on a wide variety of compute systems, downloading the LINC-processed data from dCache storage at SURF and uploading the results there. 56\% of the new fields for DR3 were processed at the University of Hertfordshire high-performance computing facility, 30\% on SURF, 11\% at Leiden University, 2\% on facilities of the UK SKA Regional Centre (UKSRC) and 1\% on the LOFAR-IT infrastructure at INAF-IRA in Bologna. A central database was used to keep track of all processing and ensure efficient distribution of the work.

\section{Individual pointing image quality}
\label{Sec:individual_pointings}

To assess the quality of individual pointings, we used the Python Blob Detector and Source Finder (\textsc{PyBDSF}; \citealt{Mohan_2015}; version 1.10.3) to derive a source catalogue and characterise the noise across each of the 2551 individual LoTSS-DR3  pointing images. Prior to cataloguing we scaled each of the images with the NRAO VLA Sky Survey (NVSS: \citealt{Condon_1998}) derived flux density scaling factors (see Sec. \ref{Sec:mosaic_fluxscale}). We then followed the procedure outlined in Sec. 3.1 of \cite{Shimwell_2025} to identify unresolved isolated sources in each of these catalogues. The unresolved sources were identified using a population of seemingly compact isolated sources (defined as sources at least 30$\arcsec$ from their nearest neighbour, having \textsc{PyBDSF}
code `S' corresponding to being fit by a single Gaussian, a major
axis below twice the synthesised beam major axis and having an integrated flux
density, $S_I$, divided by the integrated flux density uncertainty, $\sigma_{S_I}$, exceeding 3). We corrected the peak brightness ($S_P$) of these seemingly compact sources for radial smearing effects. We then characterised $\ln(S_I/S_P)$ versus $S_I/\sigma_{S_I}$ of these sources and fitted an envelope that encompasses them. We used the fitted envelope together with the original \textsc{PyBDSF} catalogue to identify many more unresolved sources in each of the fields. We then removed all `S' code sources and those that are not very isolated (nearest neighbour within 30$\arcsec$) to create our compact isolated source catalogues for each pointing.

For each field we also identified dynamic-range limited regions that are more likely to contain catalogued artefacts. We identified these following the approach in Sec. 3.4 of \cite{Shimwell_2025}. Specifically, we characterised the noise as a function of distance from the source in the DDFacet residual image around sources with $S_I>5$\,mJy that are not close to any other bright sources (more than 300$\arcsec$ from sources with $S_I>10$\,mJy. We grouped together sources of similar $S_I$ to study how the noise for a source of a particular flux density varies as a function of distance from the pointing centre. We fitted an exponential to the measured values and then using the source catalogue we could identify regions of the map that are substantially impacted by dynamic range limitations ($>5$\% noise increase). We added to our compact isolated source catalogues information about whether the sources are in a dynamic range limited area. We find that typically 7.6$\pm$3.1\% of the area within 30\% of the power primary beam of each of the images are dynamic range limited.

We also attempted to remove poorly calibrated facets. Typically the fields are calibrated in 45 directions and occasionally the calibration in one or more facets can diverge. Reasons for this include the quality of the data as well as the properties of the target field, such as dynamic range limitations or regions with low apparent flux density. In some cases the facets where the calibration has failed can be obvious and display clear noise discontinuities or bright artefacts. However, in other cases identifying them is more challenging.  We tried to identify these failed regions by creating a figure of merit for each facet in the image. This is based on the properties of a Gaussian fit to the residual image facet noise pixels (i.e. height, width, location), astrometric offsets from the Pan-STARRS DR2 optical catalogue (\citealt{Flewelling_2020}),  dynamic range, and whether discontinuities can be detected in the noise properties of pixels at the boundaries between facets. If a particular facet is an outlier in any aspect then its figure of merit decreases and those that fall below a given threshold are identified as possibly bad. Finally, we performed a visual inspection of the failed facets to indeed ensure that we were only removing genuinely failed facets from the individual pointing images. These failed facets were removed in mosaicing, as is described further below in Sec.\ \ref{Sec:mosaic_quality}. 

The LoTSS pointings are separated by 2.58$^\circ$, but even at optimal
elevations the radius of the region encompassed by the 30\% primary
beam power level is 2.75$^\circ$. Hence there is a large (up to 24.6
deg$^2$) overlap between neighbouring pointings, containing
many detectable sources (typically 200 deg$^{-2}$), which
allows us to study the variations in their properties in
the different images. We created a cross-matched catalogue for the
compact isolated sources identified in each individual pointing. To do
this we performed a nearest neighbour cross-match ($<20\arcsec$) between
the central pointing compact isolated catalogue and the catalogues of
the neighbouring pointings before any filtering is applied to them
(except that we remove sources not within 30\% of the power primary beam). We remove sources where the identified nearest neighbour is not identified as compact in the neighbouring pointing compact source catalogues. Then for all other sources we record the catalogued $S_I$, $S_P$ and associated errors in each of the catalogues. We used these \textsc{PyBDSF} noise images, catalogues of compact isolated sources, and cross-matched catalogues to assess and refine the individual pointing image quality in several ways as described in the following subsections. 

\subsection{Astrometric alignment}
\label{Sec:astrometric}

We improved the astrometric accuracy of our images by aligning them
with the Pan-STARRS DR2 optical catalogue, which has positional
accuracies within 0.05$\arcsec$ (\citealt{Magnier_2020}, following the
procedure established for LoTSS-DR1 and described by \cite{Shimwell_2019}. In summary, for each calibration facet in a LoTSS pointing, we measured astrometric offsets by finding the peak in the histograms of RA and Dec separations between the Pan-STARRS catalogue and the LoTSS individual pointing catalogue. The pixels in each calibration facet were then independently shifted in both RA and Dec to align the astrometry of the image with Pan-STARRS. The procedure is almost identical to that used for previous LoTSS data releases, but instead of using the full individual pointing LoTSS catalogue we use the compact isolated source catalogue, which reduces the influence of complex sources that are challenging to align. Furthermore, to easily facilitate refinements of the astrometric alignment procedure, we next applied the corrections during the mosaicking process (see Sec. \ref{Sec:mosaic_quality}) rather than during the final imaging with DDFacet. 

LoTSS-DR3 spans a large fraction of the northern sky and includes challenging low declination and Galactic regions. In some of these challenging cases we are unable to find a clear peak in the Pan-STARRS to LoTSS offsets due to, for example, the low number of sources or the complexity of the region. In cases where the derived astrometric offset corrections are smaller than the errors in those corrections, we choose not to apply them. For facets where offsets cannot be derived but offsets are available for other facets in the field, we apply the median of the derived astrometric offsets. If no reliable astrometric offsets were found for any facets in the field, no corrections are applied. In total only 57 fields (or 3\% of sources) have no astrometric corrections applied, but for 40\% of all facets (corresponding to 44\% of sources) we are unable to derive facet corrections and apply the median found for the field.

\begin{figure}[htbp]
   \centering
   \includegraphics[width=\linewidth]{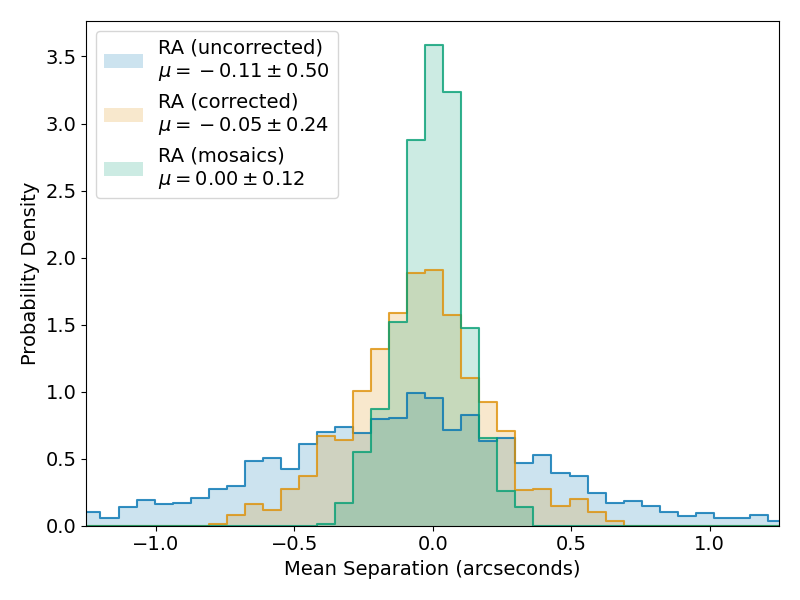}\\
   \includegraphics[width=\linewidth]{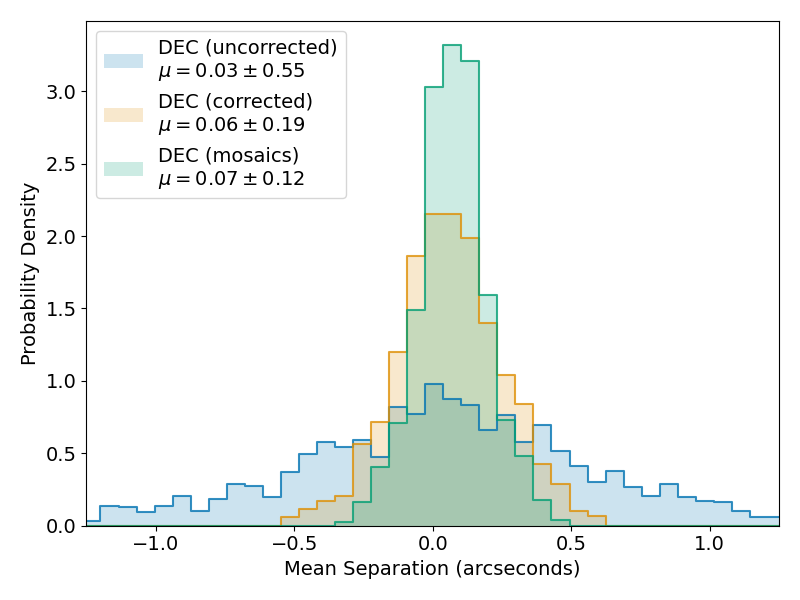}
    \vspace{-0.2cm}
   \caption{Distribution of the mean RA (top) and Dec (bottom) astrometric offsets between LoTSS-DR3 and FIRST for the different LoTSS-DR3 individual pointing and mosaic catalogues. For the individual pointing catalogues, we show the distribution of mean offsets both before and after the astrometric corrections described in Sec. \ref{Sec:astrometric} were applied. }
   \label{Fig:astrometric_alignmemt}
\end{figure}

To demonstrate the alignment procedure and assess the resulting accuracy, we cross-matched our individual pointing catalogues with the  5$\arcsec$ resolution 1.4\,GHz FIRST survey catalogue (\citealt{Becker_1995}). In Fig. \ref{Fig:astrometric_alignmemt}, we show the offsets between LoTSS-DR3 and FIRST sources, including only the 1,360 LoTSS pointings where more than 200 sources with a signal-to-noise ratio (S/N) exceeding 10 and within 30\% of the power primary beam are cross-matched (within 3$\arcsec$) with FIRST. The astrometric alignment significantly improves the typical median offset relative to FIRST and  slightly decreases the amount of scatter in the offsets. Some low level offsets still remain in the data but these are primarily at lower declination ($\sim 20^\circ$) or in regions of severe dynamic range limitation. Overall, in this region for these sources, we find that after alignment the median of the mean separations between LoTSS-DR3 and FIRST positions for sources in a given individual LoTSS pointing is $0.05\arcsec$ in RA and $0.06\arcsec$ in Dec, with corresponding standard deviations of the mean offsets per pointing of $0.24\arcsec$ and $0.19\arcsec$.

\subsection{Flux density scale refinement}
\label{Sec:flux_scale_refinement}

As is shown in Fig. 9 of \cite{Shimwell_2022}, the integrated flux density scale varies across the field of view of individual LoTSS pointings. The origin of this variation is not yet fully understood, although it is thought that errors in the current LOFAR beam model may play a role. Here we attempt to correct for these variations by deriving and applying scaling factors that vary as a function of RA and Dec across each pointing.

We measured the ratio of integrated flux densities between every pointing and its neighbouring pointings after cross-matching the individual field isolated compact-source catalogues. For each central pointing, using only sources that are cross-matched with at least one of its neighbours, we used $k$-means clustering to group sources in RA and Dec into up to 100 regions. If the least populated region contains fewer than ten sources, we reduced the number of clusters until every region contains at least ten, ensuring that representative integrated flux density ratios are obtained across all regions. The flux density ratios approximately show a linear trend with position around the field centre. Using the mean RA and Dec of each cluster, together with the median of the base-10 logarithm of the integrated flux density ratios of the cluster sources, we fitted a plane of the form 
\begin{equation}
\operatorname{median}\!\left(\log_{10} \frac{S_{\text{I,ref}}}{S_{\text{I,other}}}\right)
= a \,\Delta \mathrm{RA} + b \,\Delta \mathrm{DEC} + c
,\end{equation}
where $\Delta \mathrm{RA} = \mathrm{RA} - \mathrm{RA_{center}}$ and $\Delta \mathrm{Dec} = \mathrm{Dec} - \mathrm{Dec_{center}}$ are the right ascension and declination offsets of the sources from the pointing centre, respectively, and $S_{\mathrm{I,ref}}$ and $S_{\mathrm{I,other}}$ are the integrated flux densities from the central and neighbouring pointings. The coefficients $a$ and $b$ describe the gradients of the flux density scale across the pointing, and $c$ gives the flux density correction at the pointing centre. Since the LoTSS flux density scale is already tied to NVSS (see Sec. \ref{Sec:mosaic_quality}), we re-normalised the plane such that its mean correction factor over the cross-matched region is unity, preventing any systematic drift.

This procedure was performed independently for each pointing, producing alignment planes for all pointings in the survey. We then calculated the corrections to the catalogue of each central pointing by evaluating the plane at the RA and Dec of all sources relative to the pointing centre.  If the correction were applied to a single central pointing only, the plane would align the flux densities with neighbouring pointings to the extent that the plane fits the data. However, because each plane is derived from the integrated flux density ratios between a central pointing and its neighbours, we only applied half the derived scaling factor, assuming that each pointing contributes equally to the misalignment.

\begin{figure}[htbp]
   \centering
   \includegraphics[width=0.9\linewidth]{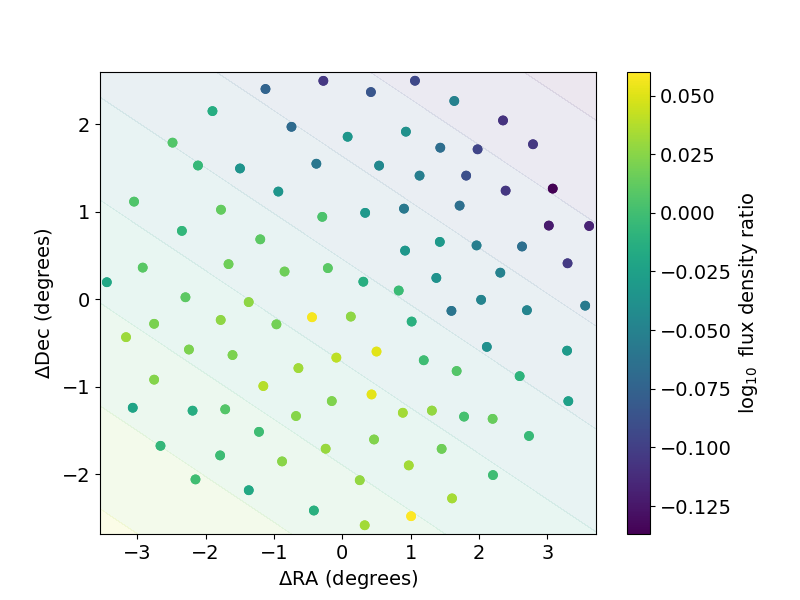}
   \includegraphics[width=0.9\linewidth]{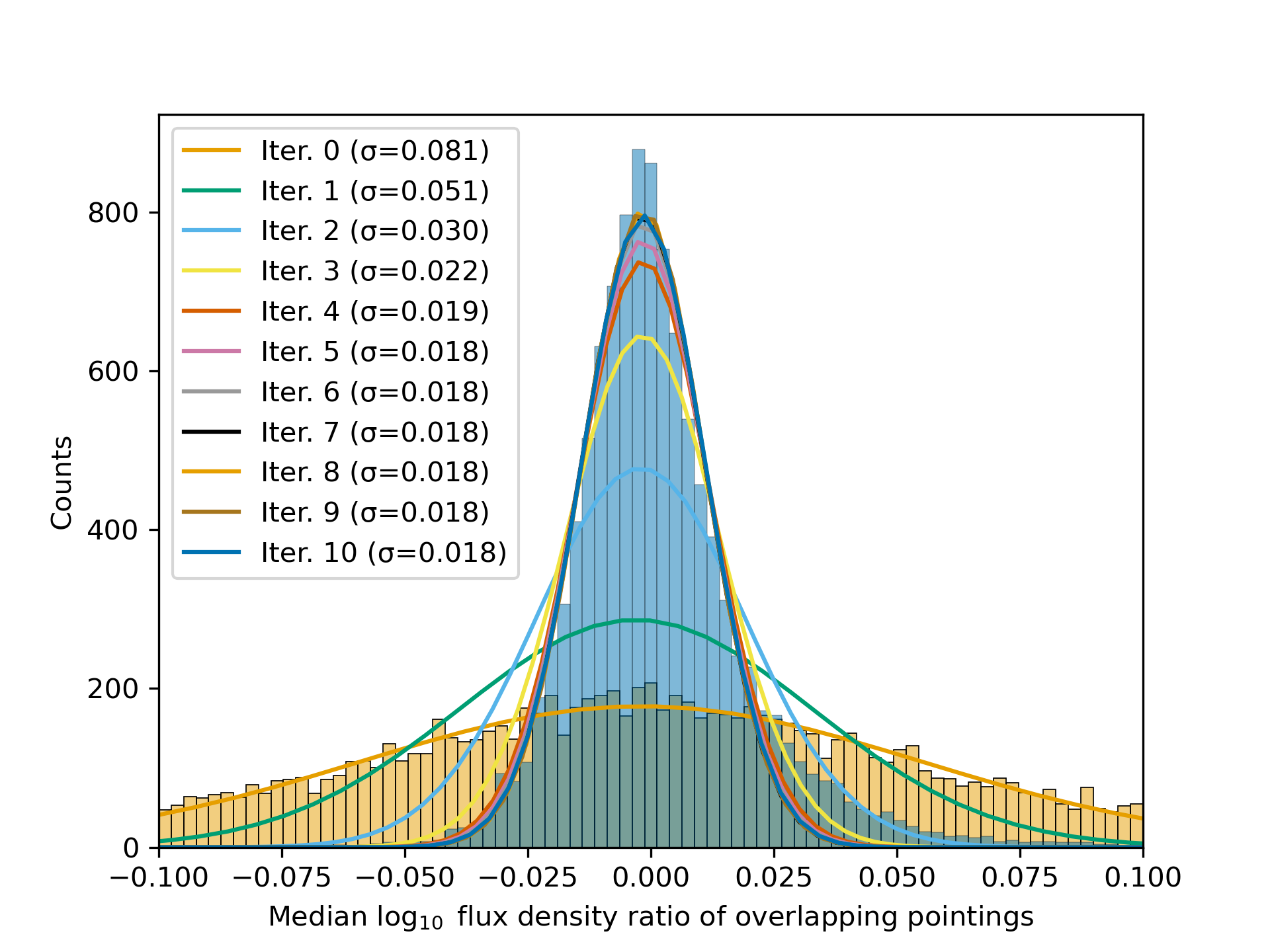}
    \vspace{-0.2cm}
   \caption{Flux density scale alignment of LoTSS-DR3. Top: Median ratio of the catalogued integrated flux density between pointing P028+41 and the neighbouring fields before any correction to the flux density scale has been applied. The colour background shows the plane fit to the integrated flux density ratios that we use to align this pointing with the neighbouring fields. The colour bar corresponds to both the points and the background. Bottom: How the distribution of the median integrated flux density scale ratio changes as different iterations of the alignment corrections are progressively applied. The standard deviations of the fitted Gaussian functions are displayed in the legend.}
   \label{Fig:fluxscale_alignmemt}
\end{figure}

Even though the same sources are used in a given overlap region between two pointings, the corrected flux density scale for that region can differ between neighbouring pointings because each plane is fit considering the overlaps with all neighbours to a particular central pointing. We therefore repeated the entire process ten times, each time using the rescaled catalogues as input. This progressively refines the alignment and the standard deviation of the base-10 logarithmic flux density scale ratios decreases from 0.081 to 0.018 (see Fig. \ref{Fig:fluxscale_alignmemt}), corresponding to a reduction in typical integrated flux density variations from 20\% to 4\%. 

After the final iteration, we produced a scaling image for a particular LoTSS pointing, which is the product of the planes derived from the ten iterations where that pointing was the central pointing. Finally, we note that 14 fields have no neighbouring pointings, and therefore no correction could be derived for them. This leaves 2,537 fields where, during the mosaicking, we applied position-dependent flux density scale correction factors.

\subsection{Image sensitivity}

\begin{figure}[htbp]
   \centering
   \includegraphics[width=0.95\linewidth]{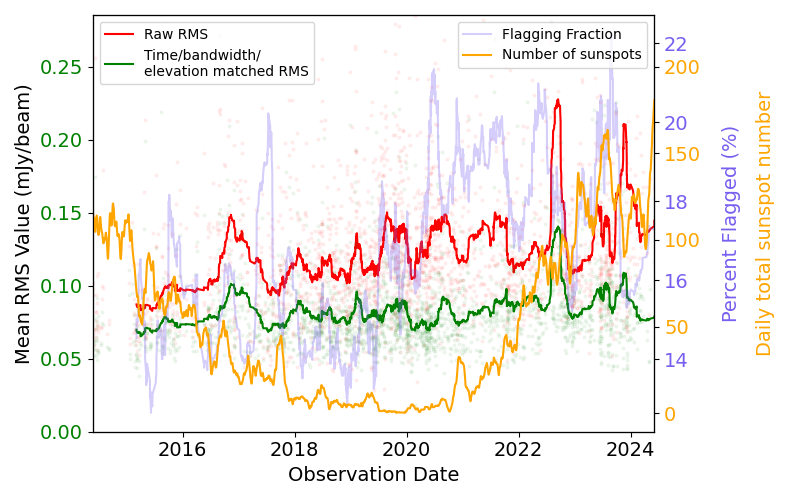}
   \vspace{-0.2cm}
   \includegraphics[width=0.95\linewidth]{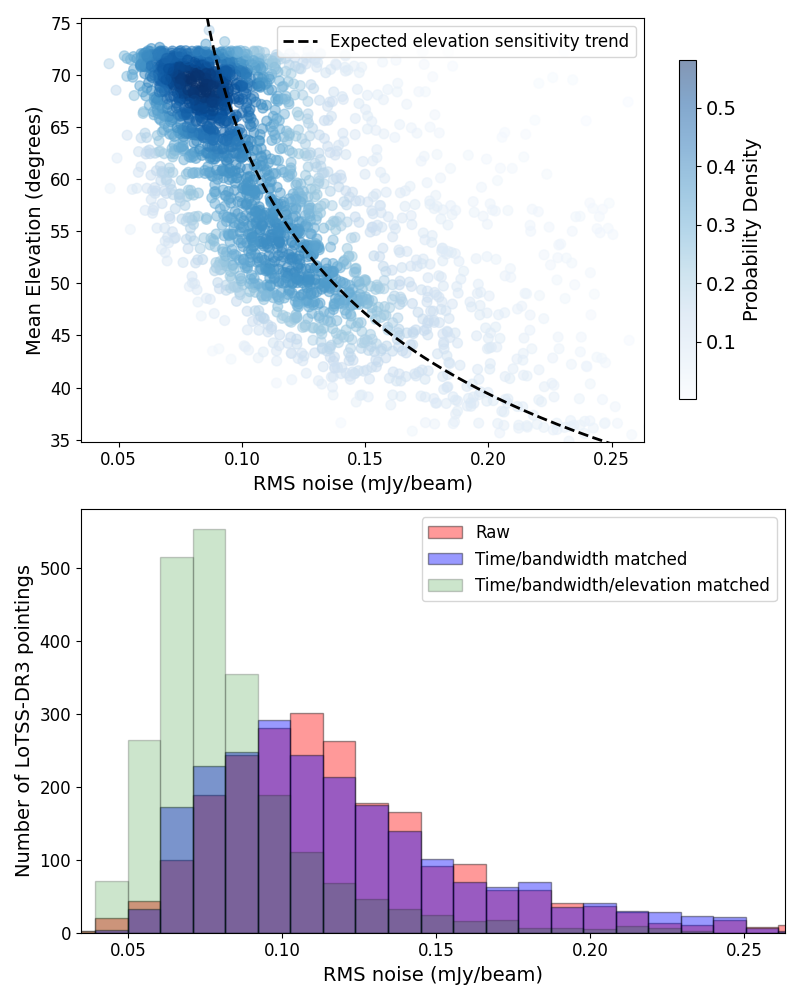}
    \vspace{-0.2cm}
   \caption{Top: Individual pointing `matched RMS'
     as a function of observation date (green) together with the raw
     RMS measured from the images. The `matched RMS' is the raw RMS
     adjusted to account for observation elevation, duration and
     bandwidth differences; it is scaled to match an 8\,hr 231 sub-band
     observation at optimal elevation ($\sigma_{Matched} =
     \sigma_{Measured} \times \cos(\pi/2 - \epsilon)^2 \times \left(
     \frac{(1-F)\times T \times N_{SB}}{8.0\times231} \right)^{0.5}$
     ), where $\epsilon$ is the elevation, $F$ is the fraction
     flagged, $T$ is the observation duration in hours, and $N_{SB}$ is
     the number of sub-bands. Also plotted is the flagging fraction and
     the daily total number of sunspots (\citealt{Clette_2015}) which
     gives an indication of the solar activity throughout the
     observing range. Middle: RMS adjusted to
     account for observation duration and bandwidth demonstrating that
     the expected elevation dependence is observed. Bottom: Histogram of the raw and matched RMS values.}
   \label{Fig:RMS_v_date}
\end{figure}

The LoTSS-DR3 observations were conducted over a 10.5\,yr period and although the data were generally collected using the same setup there are variations in the observing elevation, observation duration, and observing bandwidth. Furthermore, the observations span more than an entire solar cycle and there are significant changes in solar activity and ionospheric conditions throughout the observing period. In Fig. \ref{Fig:RMS_v_date} we show the measured RMS of the individual fields against the observation date. Here the RMS is the mean \textsc{PyBDSF} noise for sources catalogued in the field accounting for the primary beam attenuation thus providing a measure of the central image RMS. The date is the mean date if the field was observed with several different observations. We also show the matched RMS values to account for differences in the observing elevation, duration and bandwidth and demonstrate that the elevation dependence of our sensitivity matches expectations from the projected size of the LOFAR HBA antenna fields.

Excluding regions within 10$^\circ$ of the Galactic plane, the individual pointing images in LoTSS-DR3 have central RMS noise levels ranging from 29\,$\mu$Jy beam$^{-1}$ to 438\,$\mu$Jy beam$^{-1}$, with a median of 112\,$\mu$Jy beam$^{-1}$ and a standard deviation of 46\,$\mu$Jy beam$^{-1}$. Even in the deepest regions of LoTSS-DR3, the RMS is sufficiently high that the images are not severely impacted by confusion noise, which is expected to increase the RMS by at most $\sim5\%$ (see \citealt{Shimwell_2025}).

\section{Mosaicking and image quality}
\label{Sec:mosaic_quality}

Our final images were made, as for DR2, from a weighted combination of
all the data that can contribute to any given position on the sky,
where the weight takes account of the local (per facet) RMS noise
level and the telescope beam. However, we adopted a new mosaicing
strategy for DR3 in order to get the most out of the available
observations. The sky was divided into equal-area
HEALPix\footnote{\cite{Gorski_2005}: \url{https://healpix.sourceforge.io/}} pixels (we
use {\tt NSIDE=16}, which gives a pixel area of 13.4 deg$^2$) and, for each
field observation, we computed which pixels that field contributes to
above a cut-off corresponding to the 30\% level of the power primary
beam. We also masked out small boundary regions of some fields where the
pixels are affected by the subtraction of sources outside the main
field of view. Then, for each HEALPix pixel, we created a square grid of
1.5-arcsec (full resolution) or 4.5-arcsec (20-arcsec resolution)
pixels centred round the HEALPix centre and extending out to 10\%
larger than the maximum extent of the HEALPix boundary in RA or Dec,
and then do a facet-by-facet reprojection of all the fields
contributing to that HEALPix pixel onto the grids. Astrometric
positional corrections derived as discussed in Sec.
\ref{Sec:astrometric} and flux scale corrections discussed in Sec.
\ref{Sec:flux_scale_refinement} are applied at this point, and bad
facets, identified as discussed in Sec. \ref{Sec:individual_pointings},
are excluded. For full-resolution mosaics where images at positions
below declinations of $10^\circ$ contributed, we convolve all the
contributing images to a resolution of $9 \times 9$ arcsec before
regridding them (by convolving the sky model and restoring to the
residual image to mimic the effect of using the appropriate restoring
beam). During mosaicking, each individual pointing image is cut at the
30\% level of the power primary beam, and afterwards blanked mosaics
are produced by removing pixels from the square image that lie outside
the boundaries of the HEALPix pixel. The overall intention of this
strategy is to ensure that any valid pixel of a LoTSS observation
(within the 30\% level of the power primary beam) is represented in
the final mosaic images, while minimising duplication of sky area. An
image of most sky areas can be obtained by finding the area within the
mosaic corresponding to the HEALPix in which it lies. Very large
sources may be split across more than one HEALPix mosaic, but of
course we can form a mosaic using the algorithm above at any sky
position. 

Once mosaics have been created, we used \textsc{PyBDSF} to extract a
per-mosaic source catalogue for the unblanked full-resolution mosaics.
This uses the same modified \textsc{PyBDSF} strategy as described for
ELAIS-N1 by \cite{Shimwell_2025}, with a $4.5\sigma$ initial
sensitivity cut. We expect this to produce slightly more sources than
our simpler source finding in DR2, but the main effect is that
completeness is expected to be significantly increased at the faint
end of the source counts. Use of the unblanked mosaics ensures that
catalogues are not incomplete at a HEALPix pixel boundary. The
resulting per-HEALPix catalogues are then filtered to contain only
sources whose RA and Dec lie within the relevant HEALPix pixel.
Finally, we generate a combined source catalogue for the whole survey
by stacking the filtered per-mosaic catalogues, which are
non-redundant by construction.

\begin{figure*}[htbp]
   \centering
   \includegraphics[width=\linewidth]{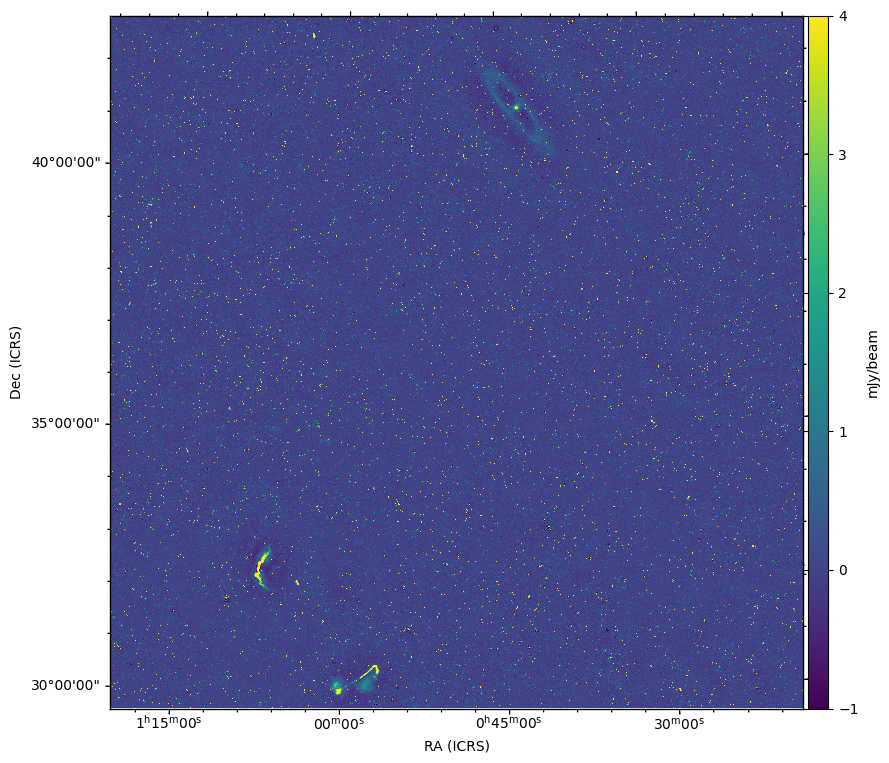} 
    \vspace{-0.2cm}
   \caption{Example 45-deg$^2$ region of the extragalactic sky from LoTSS-DR3. Such a region typically contains around 30,000 sources detected above $4.5 \times \mathrm{RMS}$. Prominent in this image are the large radio galaxies NGC 315 (bottom) and 3C 31 (lower centre left), as well as the spiral galaxy M 31 (top).}
   \label{Fig:extragalactic_image}
\end{figure*}

\begin{figure*}[htbp]
   \centering
   \includegraphics[width=\linewidth]{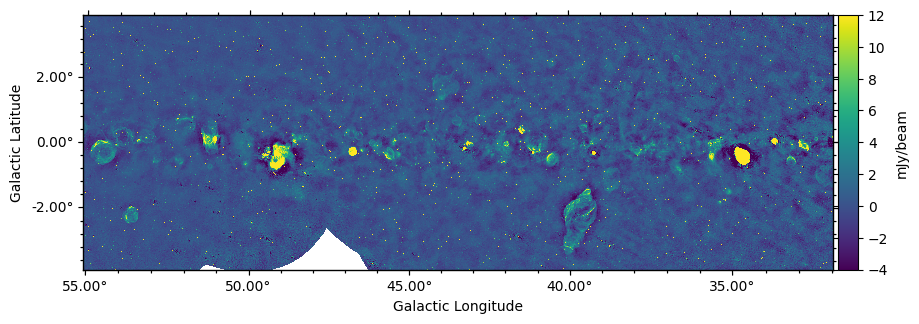} 
    \vspace{-0.2cm}
   \caption{Region of the Galactic plane with the highest density of known supernova remnants in LoTSS-DR3. Spanning 190 deg$^2$, it is centred at a Galactic longitude of $43.5^\circ$ and a latitude of $0.0^\circ$. The area includes several prominent supernova remnants, such as G054.4-00.3, G049.2-00.7, G043.9+01.6, G039.7-02.0, and G034.7-00.4. The blanked pixels are outside the coverage of LoTSS-DR3.}  
   \label{Fig:galactic_image}
\end{figure*}

\begin{figure}
\includegraphics[width=\linewidth]{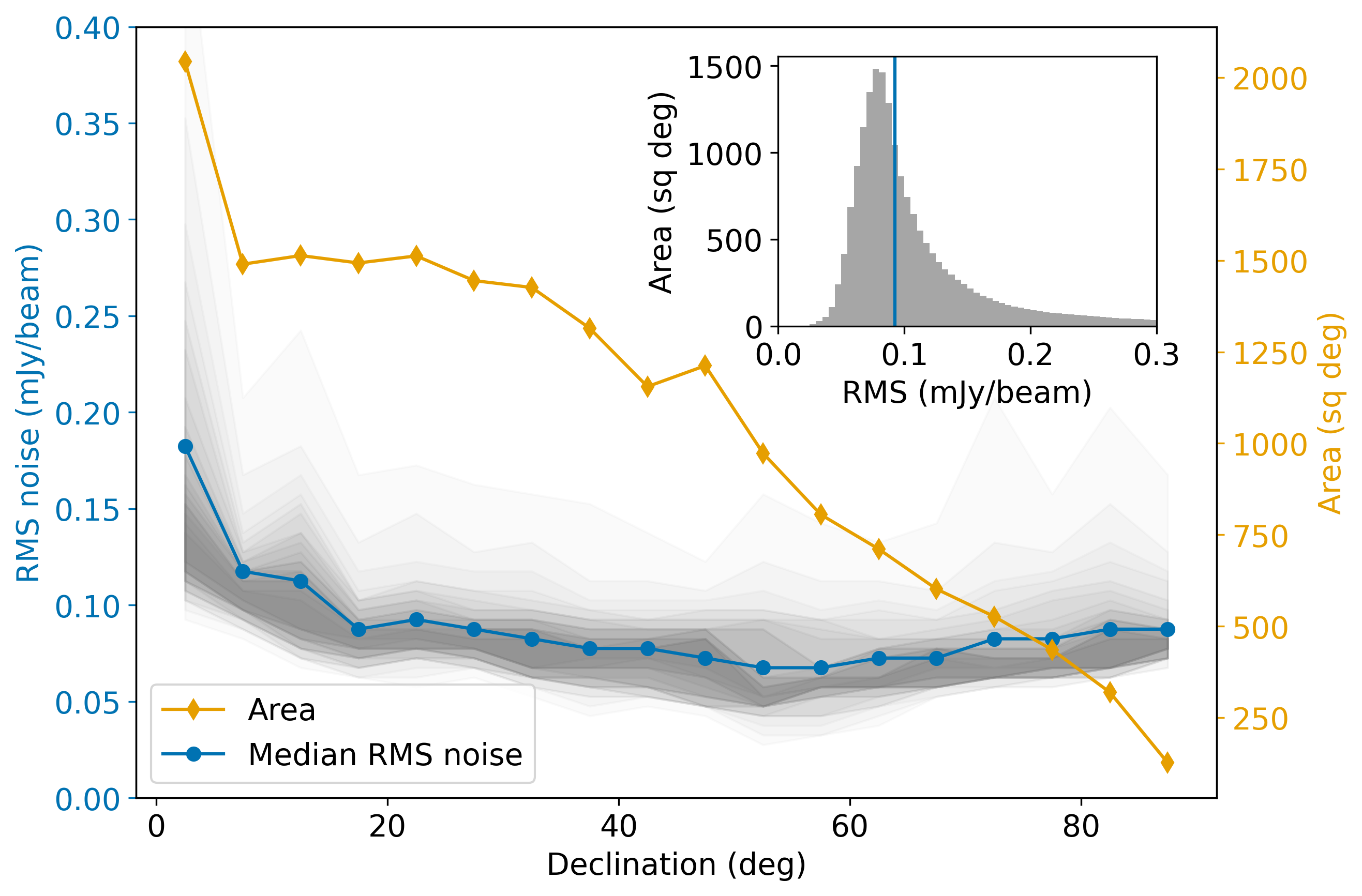}
\caption{Sensitivity of the LoTSS-DR3 mosaics as a function of declination. The blue points and line show the median sensitivity as a function of declination in 5$^\circ$ strips. The orange curve shows the sky area covered by our mosaics in each declination strip. The grey shading shows the 95\%, 90\%, 85\%  etc sensitivity level for the noise pixels in each declination range. The inset shows the distribution of noise in the entirety of LoTSS-DR3 and the median noise level of 92$\mu$Jy beam$^{-1}$ is marked in blue.}
\label{Fig:mosaic_sensitivity}
\end{figure}

Typical extragalactic and Galactic images for LoTSS-DR3 are presented in Fig. \ref{Fig:extragalactic_image} and Fig. \ref{Fig:galactic_image}. 
In Fig. \ref{Fig:allsky_projection} we show the local RMS of LoTSS across
the sky where the values plotted are those recorded in our mosaicked
\textsc{PyBDSF} noise images for the mosaics, regridded for convenience of
plotting onto a HEALPix grid with pixel area 0.013 deg$^2$. In Fig.
\ref{Fig:mosaic_sensitivity} we show the same sensitivity measurements
but  as a function of declination, as well as the noise distribution
of the entire surveyed region. In total the 1,580 mosaics made available
cover 19,035 deg$^2$, or 46.1\% of the whole sky.

\subsection{Flux density scale accuracy}
\label{Sec:mosaic_fluxscale}

In this section we consider the flux density scale of the resulting
mosaics. The flux density scale of DR3 is adjusted, as was that of
DR2, using a method based on the ratio of the flux densities of
isolated sources to those in the NVSS as discussed by
\cite{Hardcastle_2021} to derive per-field correction factors which
are applied during mosaicing. The NVSS flux density scale was tied to
that at 144 MHz using matches to the 6C survey \citep{Hales_1990} over
2,000 deg$^2$ of DR2 between $130^\circ < \mathrm{RA} < 250^\circ$,
$40^\circ < \mathrm{Dec} < 65^\circ$ In principle, 6C by construction is on
the scale of \cite{Roger_1973} since it includes calibration scans on
Cygnus A, and it was carried out at a central frequency of 151 MHz,
very close to the 144 MHz used by LoTSS. The match between bright
sources in 6C and NVSS gives us a typical spectral index $\alpha =
-0.78$, which we used to scale NVSS flux densities to 144 MHz in what
follows. Previous analyses by \cite{Hardcastle_2021} and
  \cite{Shimwell_2022} have estimated the systematic flux density
  scale offset associated with this procedure to be within 2\%, and
  here we focus on estimating the random flux density scale error.

\begin{figure*}
\includegraphics[width=0.495\linewidth]{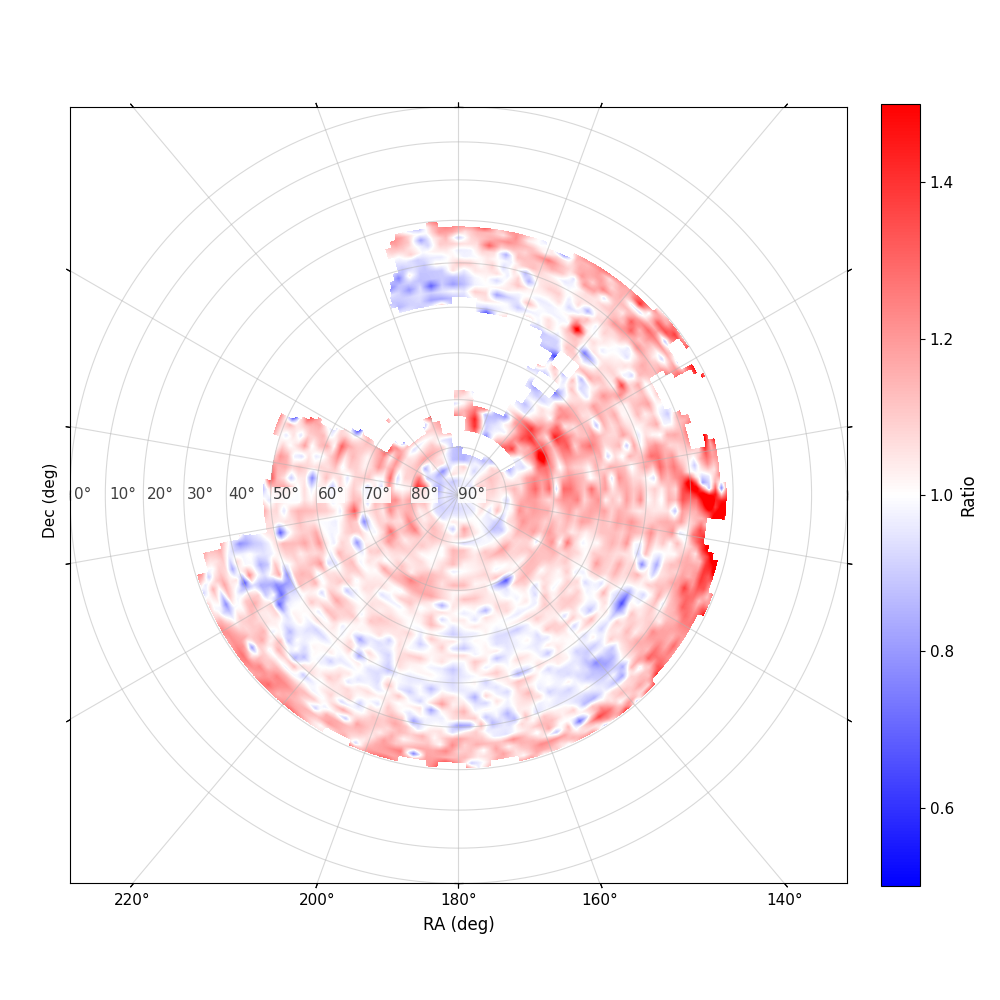}
\hspace{0.01\linewidth}%
\includegraphics[width=0.495\linewidth]{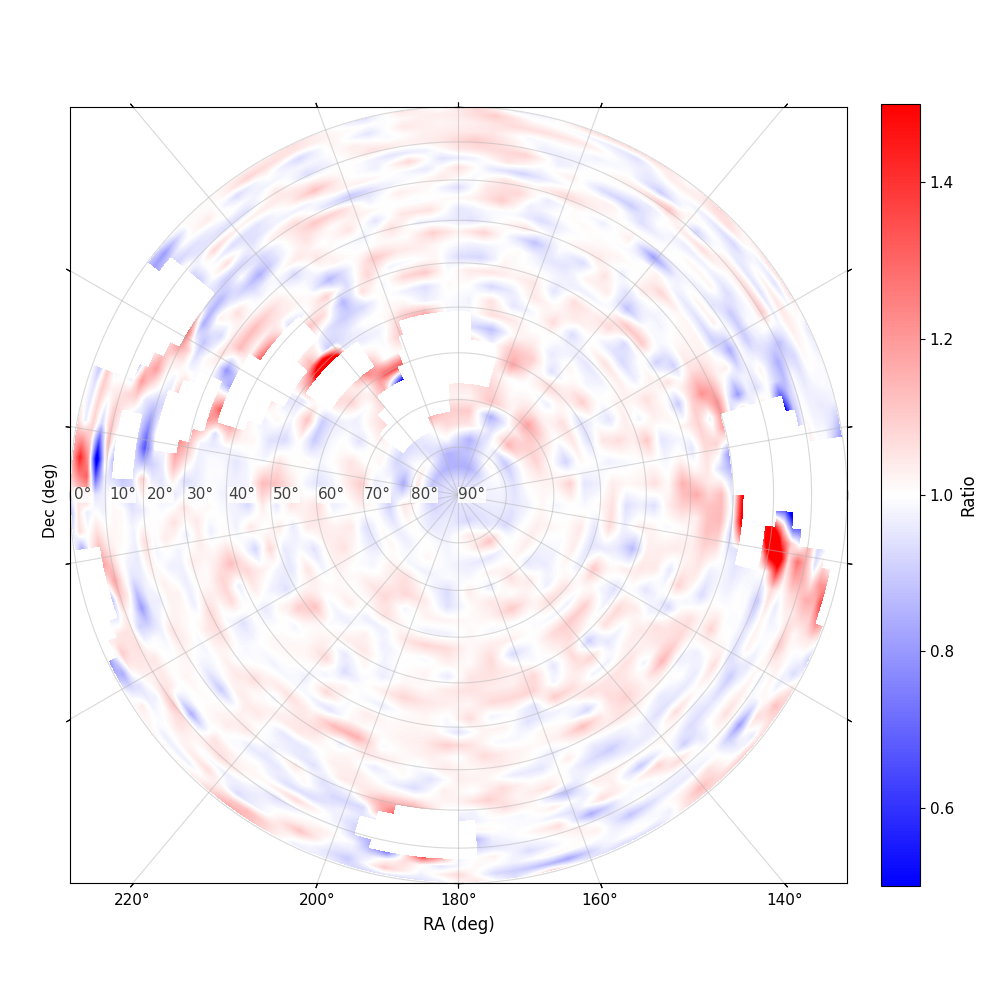}
\vspace{-1cm}
\caption{Ratio between DR3 flux density and comparison surveys: (left)
  6C and (right) NVSS with the default flux density scale. On both
  plots the colour scale is linear. 6C shows systematic strong
  deviations at lower declinations, while the alignment with NVSS is
  good except near the Galactic plane where there are small numbers of
matching sources.}
\label{Fig:NVSS6C}
\end{figure*}

To compare DR3 with 6C and NVSS, we impose lower flux density cuts on
the integrated flux density of DR3 sources at, respectively, 100 mJy
for 6C and 40 mJy for NVSS, and then filter both surveys to include
only isolated sources (with no nearest neighbour within 2 arcmin). The
flux density cut for NVSS is set high because below this level the
distribution of LoTSS-NVSS spectral indices is substantially affected
by the level of the cut \citep{Hardcastle_2016}. Sources are then
matched between DR3 and the comparison survey using a
maximum-likelihood crossmatch method that includes a likelihood term
for the flux density as described by \cite{Heald_2015} and
\cite{Hardcastle_2016}. We segment the sky covered by the 6C and
NVSS catalogues into HEALPix pixels matching the mosaics and find the median flux density
ratio over all matched sources in each pixel in order to look for any variations as a function of position. Results of this comparison are shown in Fig.\ \ref{Fig:NVSS6C}. It can
be seen that there is significant variation as a function of position
in 6C (which also appears in a comparison with the 7C survey), with
changes of flux ratio around 30\% across the sky, but
much less in NVSS. The flux density ratios are close to unity in 6C
for the region where we originally matched them, but tend to deviate
to higher values outside, particularly at low declination for 6C
($\mathrm{Dec} \approx 30^\circ$) and close to the Galactic plane. Because
these spatial variations in the flux density ratio are not present for
NVSS (as of course they should not be given that our flux density
scale derives from scaling to NVSS per field) and because they lie in regions that are
special to 6C, we conclude that they are probably related to the
calibration of 6C rather than problems with the DR3 flux density
calibration. Extreme pixels in the NVSS comparison lie around the
Galactic plane, where the number of sources corresponding to a given
HEALPix pixel are small. The standard deviation of the median LoTSS-NVSS ratios about the mean (1.01) is 0.06 for the 1,278 HEALPix pixels that
contain more than 100 matched sources, and this can be taken as an
estimate of the remaining uncertainty on the LoTSS flux density scale
from this method, bearing in mind that we would expect the flux
density scale in NVSS to be good to the $\sim 3$ -- 5\% level\footnote{\url{https://science.nrao.edu/facilities/vla/docs/manuals/oss/performance/fdscale}}.

\begin{figure*}
\begin{minipage}[t]{0.5\linewidth}
  \vspace{20pt} 
  \includegraphics[width=\linewidth]{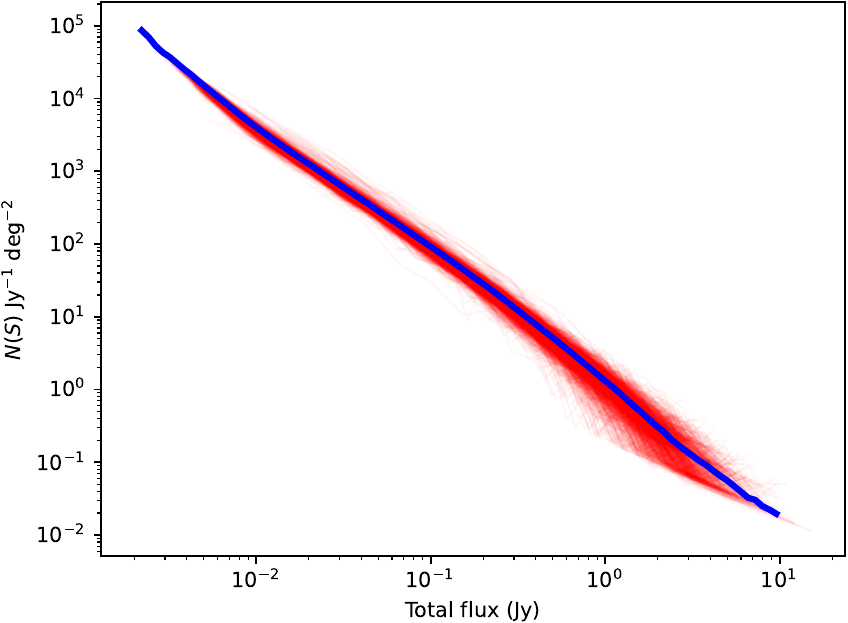}
\end{minipage}%
\begin{minipage}[t]{0.5\linewidth}
  \vspace{-20pt}
  \includegraphics[width=\linewidth]{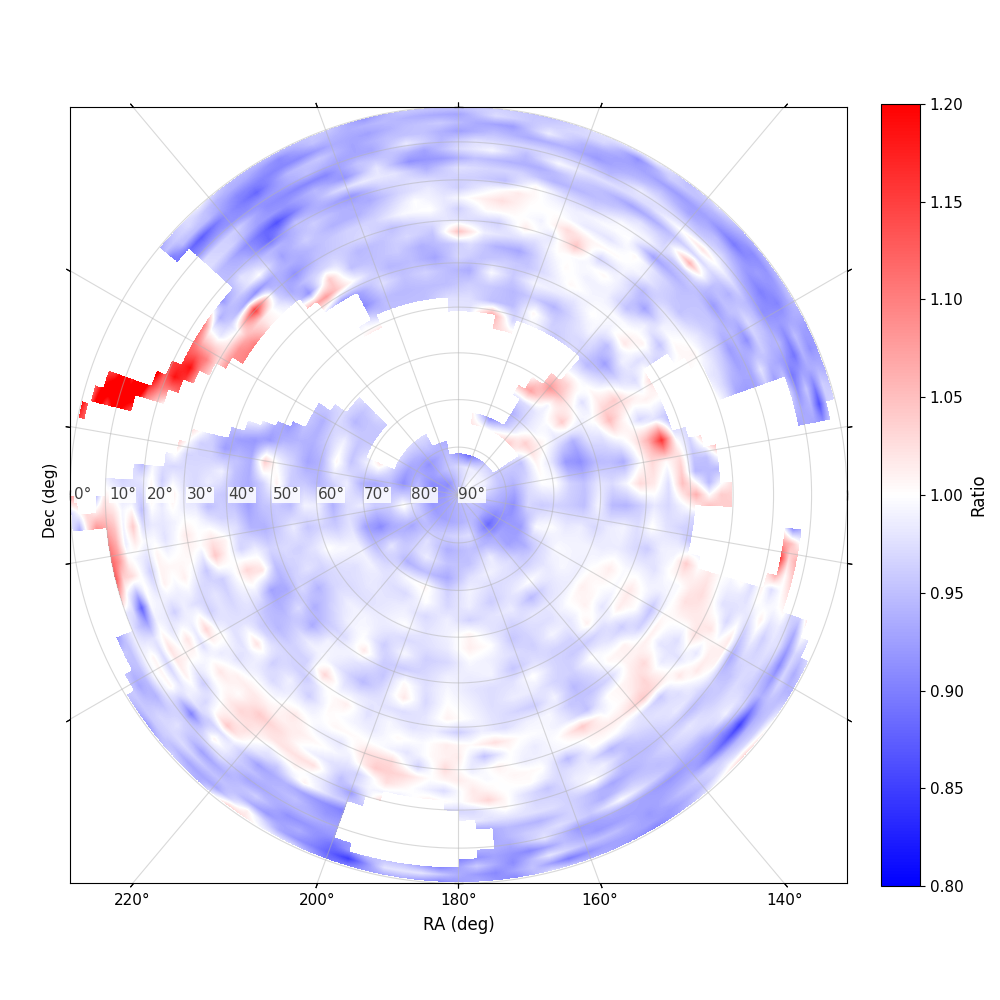}
\end{minipage}
\vspace{-0.8cm} \caption{Left: Source counts for 1413 distinct HEALPix pixels with sufficient sky area for the analysis (red) with the median source counts for the whole survey overplotted (blue). Right: Correction factor for each pixel as a function of position on the sky, where blue indicates a source count plot for that region that lies below the median (lower flux density scale or fewer sources), and red indicates one that lies above it (higher flux density scale or more sources). On both plots the colour scale is linear.}
\label{Fig:scflux}
\end{figure*}

Another way of investigating the ‘relative’ flux density scale
of the catalogue is by constructing source counts as a function of
position, since the bright end of the source counts in the field is
well constrained \citep{Intema_2017,Hardcastle_2021}. To do this we
divided the sky into independent HEALPix pixels with {\tt NSIDE=16} as
above, and constructed source counts in units of sources Jy$^{-1}$
deg$^{-2}$ for each one where more than half of the pixel contained
LoTSS sky, taking account of the area of each pixel and cutting the
source counts at 50 times the local RMS to avoid effects due to
incompleteness. We then constructed a well sampled median source count
plot from the combination of all the individual pixels. Then, for each
individual pixel, we can ask what correction to the source counts
would make the counts from that pixel best match the line. We use a
$\chi^2$ likelihood function with Poisson errors on the individual
bins. The overall distribution of these correction factors is
  quite narrow ($1\sigma$ is 3.5\%) and, as this factor includes both
flux density scale errors and cosmic variance, the implication is that
the overall flux density scale is quite consistent. However, there are
non-negligible variations over the sky, particularly in the sense that
there are fewer sources/a lower flux density scale at low declination
(see Fig. \ref{Fig:scflux}). We believe the most significant
  variations are an effect of more severe incompleteness than we have
accounted for in these regions, which often have artefacts due to
incompletely subtracted sources far down the elongated primary beam.
We also see a strong excess of sources (or higher flux density scale)
around the Galactic plane, which is unsurprising as there are many
resolved structures in these regions that will be decomposed into
sources by \textsc{PyBDSF}. Overall from these analyses we can
  conclude that the flux density scale is reasonably uniform for the
  bright sources (matched to NVSS) and has relatively small dispersion
  on scales of individual mosaics, no more than 6\% overall, but that
  there are some spatial variations in the faint source population
  that dominates the source count fits, as shown in
  Fig.\ \ref{Fig:scflux}. We do not apply any further correction to
  the flux scale as a result of this analysis.

To better understand the flux density scale accuracy and the impact of the applied refinements, we simulated the flux density scale correction procedure. We used the Tiered Radio Extragalactic Continuum Simulation (T-RECS; \citealt{Bonaldi_2019}) to create idealised 144\,MHz catalogues for sources brighter than 1\,mJy in the northern sky and from these we generated simulated LoTSS individual pointing catalogues, with sources in overlapping regions appearing in multiple catalogues. The source flux densities were  then corrupted with Gaussian noise drawn from the local RMS noise at the source position in the corresponding real LoTSS individual pointing. For each catalogue, we applied a random systematic flux density scale offset drawn from the distribution measured between the real NVSS and LoTSS individual pointing catalogues prior to correction (centred on unity with an approximately 25\% standard deviation). We imposed a planar flux density scale gradient across the field of view, allowing up to 30\% variation at the field edges while keeping the mean flux density scale fixed. Finally, we examined the residuals from the real flux density scale gradients derived in Sect. \ref{Sec:flux_scale_refinement} for each pointing and added smoothed versions of these residuals to the flux densities in the corresponding simulated LoTSS individual pointing catalogues.

We also simulated an NVSS catalogue using the same T-RECS simulations, which provide both 144\,MHz and 1.4\,GHz source flux densities based on realistic spectral-index distributions that vary with flux density. We accounted for the properties of the NVSS survey, namely that it is formed from 217,446 snapshot images, each extending 26\,arcmin from the field centre. We assume that each NVSS snapshot image has equal sensitivity (0.45\,mJy beam$^{-1}$) but include a random multiplicative systematic flux density scale error drawn from a normal distribution with a standard deviation of 5\%. The individual snapshot catalogues are combined into a mosaic catalogue, where the flux density of sources that appear in multiple snapshots is calculated using an inverse-variance-weighted mean, with the variance reflecting the primary-beam attenuation of the contributing pointings at the source location.

We then attempted to correct the simulated corrupted LoTSS individual pointing catalogues  using the same approach employed for LoTSS-DR3. To do this, we aligned each individual pointing catalogue with the simulated NVSS catalogue. Here we selected the bright (>30\,mJy) sources in the distorted simulated LoTSS image and compare these with the simulated NVSS mosaic catalogue flux densities. We first computed the global median LoTSS/NVSS flux-density ratio for the bright matched sources over all LoTSS pointings. Each individual LoTSS pointing was then multiplied by a scaling factor chosen such that the median LoTSS/NVSS ratio for that pointing  matches the global median. During this procedure, we limited the number of cross-matched sources to 310$\pm$50, which replicates the numbers used for the alignment in the real datasets (where extended sources and those with nearby neighbours are excluded). We then performed the iterative procedure described in Sect. \ref{Sec:flux_scale_refinement} to fit for the planar flux-density scale gradients across the field. Finally, we constructed a mosaiced catalogue using inverse-variance weighting, accounting for both the real primary beam shapes as well as the real spatial variations in the LoTSS individual pointing RMS levels, to combine the flux densities of sources that appear in multiple individual pointing catalogues.

\begin{figure}
\includegraphics[width=\linewidth]{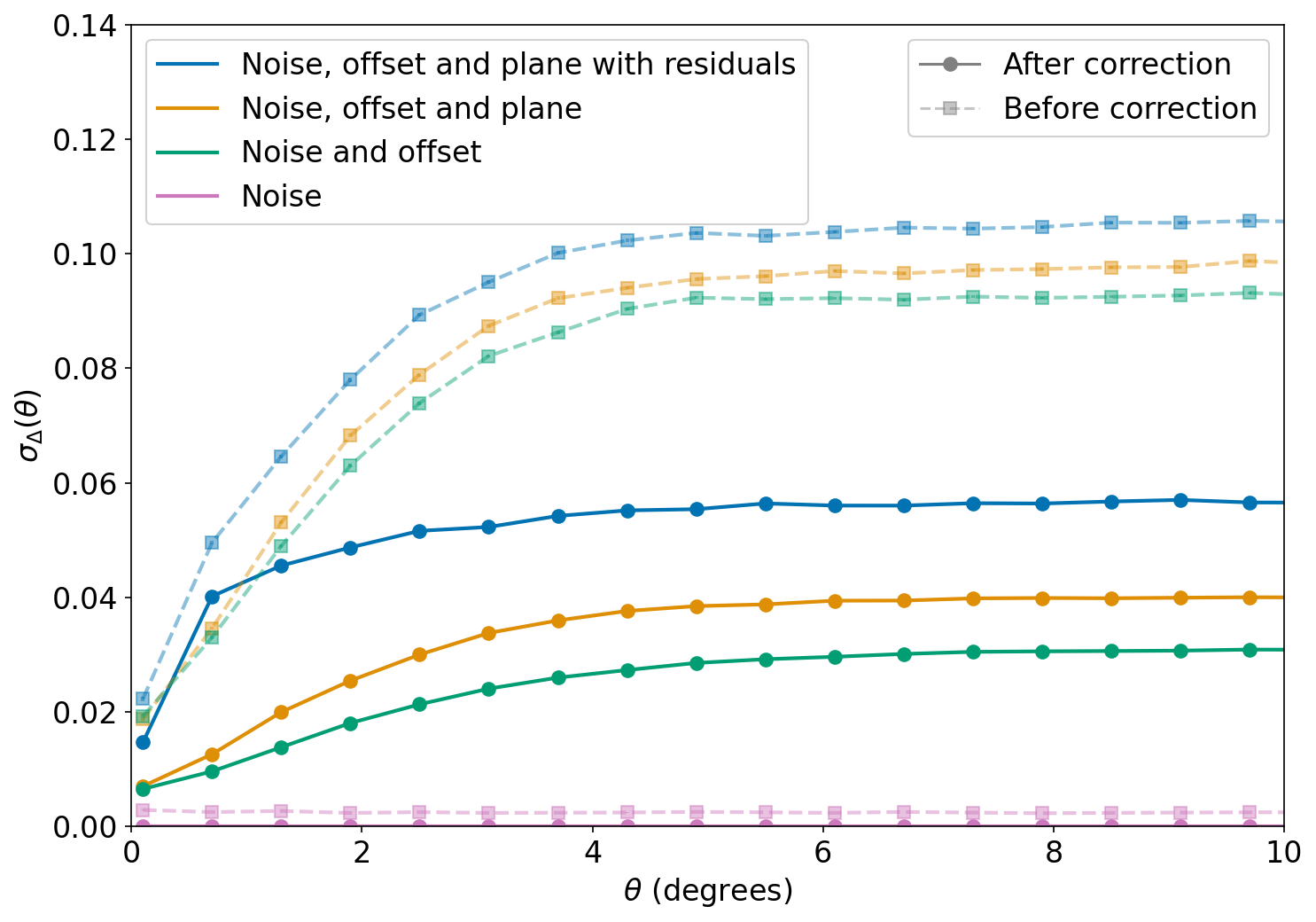}
\caption{Simulated standard deviation of the pairwise fractional difference, $\sigma_\Delta(\theta)$, between bright (>40\,mJy) sources separated by different angular scales on the sky, $\theta$. The simulations take the true flux and progressively distort it independently for each pointing with realistic thermal noise, a flux density scale offset, a planar flux density scale gradient and finally residual offsets from the gradient. Here we show the simulations with different distortions both before and after the corresponding corrections are applied to the simulated mosaic catalogues, demonstrating the impact of each corruption and the effectiveness of the refinement procedure. At large separations, $\sigma_\Delta(\theta)$ converges to 0.06 in the fully distorted and then corrected case (compared to 0.11 before any correction), whereas on small angular scales it decreases to 0.02.}
\label{Fig:pairwise_std}
\end{figure}

In exactly the same way as was done with the real LoTSS-DR3 catalogue, we used the simulated corrupted and the corrected LoTSS mosaic catalogue to measure the standard deviation of the median ratio of simulated LoTSS to  NVSS flux densities, as well as the LoTSS source counts across HEALPix pixels. We find a standard deviation of the median ratios of 0.06 and a source-count correction factor distribution width of 2.5\%,  both of which approximately match the values found in our real dataset and give us confidence that the simulations have adequately incorporated the major features and uncertainties of the flux density scale alignment process. Under this assumption, we used the simulations to probe the flux density scale in more detail by comparing the flux densities in the simulated corrupted and then corrected LoTSS mosaic catalogue with the true 144\,MHz simulated flux densities before any corruptions. In Fig. \ref{Fig:pairwise_std} we show the standard deviation of the pairwise fractional difference between bright sources (>40\,mJy) separated by different angular scales on the sky, $\sigma_\Delta(\theta)$, where
\begin{equation}
\Delta = 
\frac{S_{1,\mathrm{true}}/S_{1,\mathrm{sim}} - S_{2,\mathrm{true}}/S_{2,\mathrm{sim}}}
{0.5 \left( S_{1,\mathrm{true}}/S_{1,\mathrm{sim}} + S_{2,\mathrm{true}}/S_{2,\mathrm{sim}} \right)} .
\label{eqn:pairwise}
\end{equation}
Here $S_{\mathrm{true}}$ and $S_{\mathrm{sim}}$ denote the true and simulated (corrupted and then corrected) flux densities, respectively, for sources 1 and 2 separated by an angular distance, $\theta$. The plot shows the expectations from the full simulations and also separates the errors associated with the different contributions. Overall, in agreement with our analysis of the real data, we conclude that the flux density scale is accurate to 6\% due to the refinements we have conducted. This uncertainty decreases for sources close to each other, which we expect to have better agreement in flux density scale.

\subsection{Astrometric accuracy}

As is described in Sec. \ref{Sec:astrometric}, we refined the astrometry during mosaicking by shifting individual calibration facets in RA and Dec. The resulting positional accuracy was assessed by cross-matching with FIRST. Figure \ref{Fig:astrometric_alignmemt} shows the median RA and Dec offsets between LoTSS-DR3 and FIRST positions for sources with LoTSS $S_I/\sigma_{S_I} > 10$ in the 1,242 LoTSS-DR3 pointings (of 2,551) overlapping FIRST, using the cross-matching procedure described in Sec. \ref{Sec:astrometric}. The source positions in the two surveys agree well: the median of the mean offsets per mosaic is 0.004$\arcsec$ in RA and 0.08$\arcsec$ in Dec, with standard deviations of the mean offsets per mosaic being $0.12\arcsec$. The positional accuracy in the mosaics is higher than in the individual pointings because sources are detected with higher significance in the combined images.

\begin{figure}
\includegraphics[width=\linewidth]{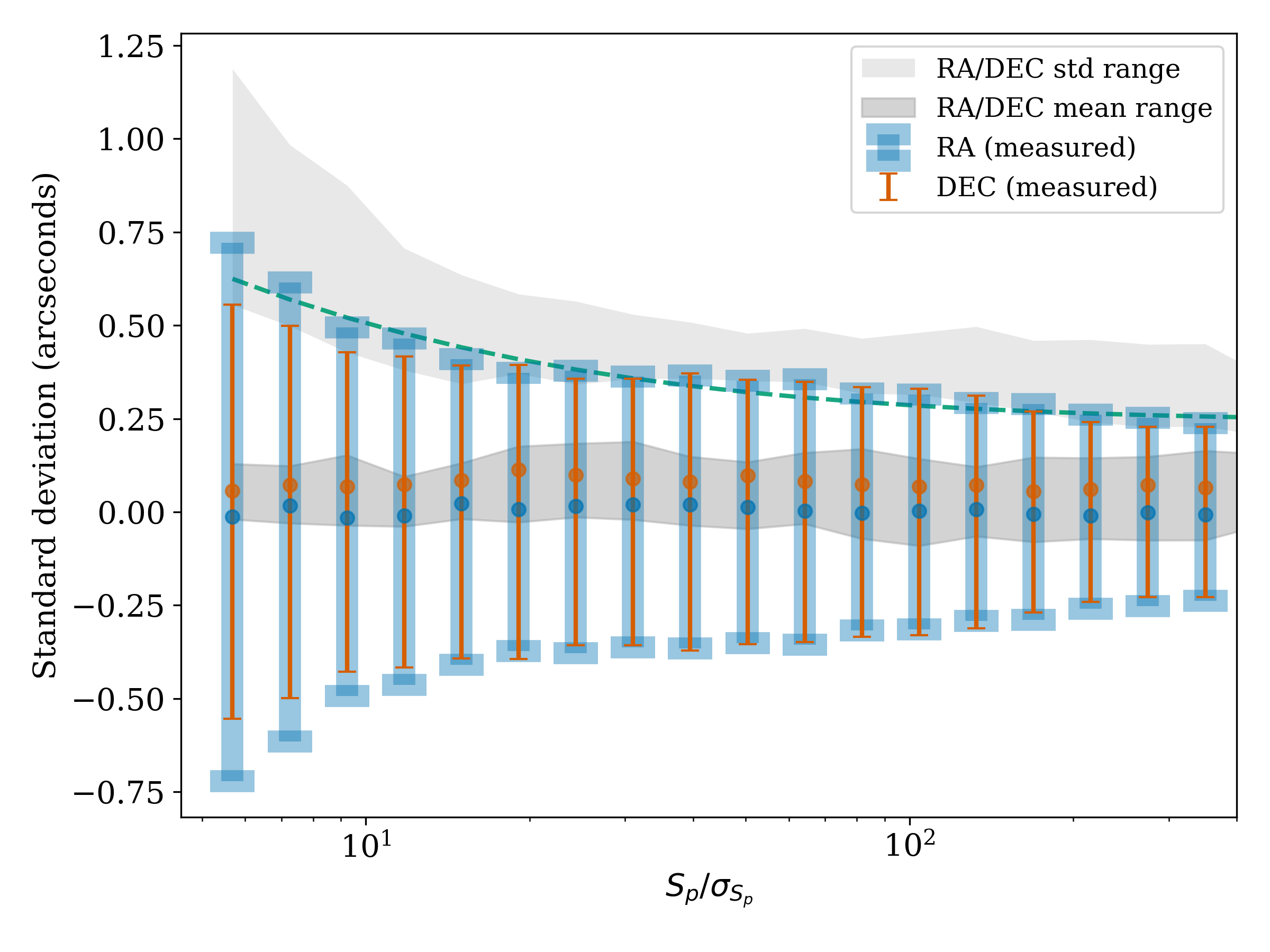}
\caption{Mean offsets (points) and corresponding standard deviations (error bars, showing $\pm 1 \sigma$ centred on zero for clarity) of RA (blue) and Dec (orange) between cross-matched LoTSS-DR3 mosaics and FIRST sources are shown as a function of LoTSS-DR3 $S_P/\sigma_{S_P}$ for the declination range 40--50$^\circ$. The green line shows the ideal standard deviations derived from modelled noise-like offsets in FIRST and LoTSS positions, including an additional systematic offset of $0.24\arcsec$. The shaded light grey region shows the range of standard deviations across the full declination range of the cross-matched catalogue ($-2^\circ$--$65^\circ$), while the darker grey region shows the corresponding range of mean offsets.}
\label{Fig:mosaic_astrometric_SNR}
\end{figure}

Figure \ref{Fig:mosaic_astrometric_SNR} shows how the mean and standard deviation of the offsets vary with $S_P/\sigma_{S_p}$ and across the declination range covered by the cross-matched catalogues. These offsets are a combination of positional errors in both FIRST and LoTSS-DR3. In the ideal case, \cite{Condon_1997} give the noise-limited 1$\sigma$ positional uncertainty in RA or Dec as $\sigma_p = \frac{1}{2} \sigma{_{rms}} \frac{\theta}{S_{p}}$,  where $\sigma_p$ is the 1-sigma positional uncertainty, $\sigma_{rms}$ is the image noise, $\theta$ is the resolution and $S_P$ is the peak brightness. This relation can be reproduced through simple tests in which perfect point-like sources are injected into our residual images and detected with \textsc{PyBDSF}, but not if any blurring of the sources is included.

To model the expected positional errors in both surveys, we assume that they reach the noise-limited uncertainty. Using the cross-matched catalogue, and taking the LoTSS position as the reference, we assign each source a random positional offset in both LoTSS and FIRST drawn from a normal distribution with the 1$\sigma$ noise-limited uncertainty, calculated from $\sigma_{rms}/S_{P}$ at the source location in each survey. We then compute the positional differences for this modelled cross-matched catalogue and group sources into S/N bins based on their LoTSS peak brightness. 

In Fig. \ref{Fig:mosaic_astrometric_SNR} we compare the standard deviation of offsets in each S/N bin for the modelled and observed cross-matched catalogues. The model reproduces the overall trend but systematically underpredict the observed standard deviations. Adding a systematic error term in quadrature aligns the modelled and observed offsets more closely. This systematic term likely arises from residual calibration errors and smearing in LoTSS, both of which cause source blurring.

Assuming all systematic errors come from LoTSS, the positional uncertainty in arcsec in RA or Dec for any LoTSS source becomes
\begin{equation}
    \sigma_p = \sqrt{ \left(\frac{\sigma{_{rms}} \theta}{2S_{p}}\right)^2 + \sigma_{sys}^2 }
,\end{equation} where the resolution $\theta$ is 6$\arcsec$ except for mosaics containing data from below Dec 10$^\circ$ where it is 9$\arcsec$. We fit for the $\sigma_{sys}$ that best aligns the distribution of the observed and modelled offsets in the cross-matched catalogues in declination strips of 5$^\circ$ and find

\begin{equation}
\sigma_{\mathrm{sys}}(\mathrm{Dec})\,[\mathrm{arcsec}] = 1.0 \times 10^{-4}\,(\mathrm{Dec} - 42.14)^{2} + 0.2448.
\end{equation}

\subsection{Compact-source completeness}
\label{Sec:completeness}

We measured the compact source detection fraction and completeness of the LoTSS-DR3 mosaics as a function of flux density. We did this by injecting artificial source populations into the \textsc{PyBDSF} residual mosaic images to preserve the spatially varying noise characteristics of the images. We injected both perfect point sources as well as more realistic source models derived from the \textsc{clean} components of actual sources in the field. In all simulations, sources with integrated flux densities above 50$\mu$Jy are injected into random positions of the mosaic. Their flux densities are drawn from differential source counts derived from T-RECS, and the 50$\mu$Jy limit was chosen to reduce the number of injections whilst ensuring that sources are injected well below the detection limits of even the most sensitive mosaics.

The simulations were performed by first constructing a simulated sky model image of the mosaic. In the model, a perfect point source corresponds to a delta-function. For realistic source models, we extracted $30\arcsec\times30\arcsec$ cutouts from the individual pointing DDFacet \textsc{clean} component model images of isolated LoTSS-DR3 sources. In both cases, we accounted for time and bandwidth smearing effects and astrometric errors in the individual pointing images that contribute to a given mosaic (typically seven). For the realistic models, we also accounted for variations in the calibration quality as these are reflected in the appearance of sources (and thus the \textsc{clean} component models) in the individual pointing images. These effects all reduce the peak brightness of sources by varying amounts across the individual pointing field of view, although the integrated flux density is conserved if the full extent of the source can be recovered. However, as source detection by \textsc{PyBDSF} is based on peak brightness thresholds, these effects impact the survey completeness.

Given a randomly assigned location and integrated flux density drawn from the T-RECS source counts, we first determined which individual pointings contribute to the mosaic image of that source. If performing simulations using the realistic source models then for each contributing pointing, we selected an appropriate source from the LoTSS-DR3 catalogue of that individual pointing. This source was chosen to closely match both the integrated flux density and separation from the pointing centre to the desired source's integrated flux density and location (we only select sources detected at a significance exceeding 20 to ensure that the \textsc{clean} component model adequately represents the source). Finally, we scaled the selected source \textsc{clean} component model so that it exactly matches the intended integrated flux density. If performing perfect point source simulations, we simply used a delta function with an amplitude equal to the desired source flux density. 

We then attempted to account for the effects of time and bandwidth smearing using the formulas from \cite{Bridle_1989}. This is necessary because there is substantial smearing in the individual pointing LoTSS-DR3 images. For example, at a distance of 2$^\circ$, the time and bandwidth smearing reduce the peak brightness of a compact source by factors of 0.89 and 0.87, respectively. During LoTSS-DR3 imaging, these smearing effects are taken into account by DDFacet by using a different point spread function (PSF) for each facet. This implies that if a source is deconvolved, it has a peak brightness that is (at least partially) corrected for smearing losses. However, sources that are not deconvolved are not corrected for such losses and subsequently have a reduced peak brightness. To mimic this behaviour in our simulations we first assessed, for every contributing pointing for a simulated source, whether the source would be deconvolved. We defined a source as deconvolved if its simulated smeared peak brightness exceeded the \textsc{clean} threshold of five times the local RMS. If the source was classed as deconvolved then we used the model images without applying smearing to replicate the DDFacet smearing correction. If it was not deconvolved we applied the total smearing corrections to reduce the peak brightness of the simulated source.  

Astrometric errors were applied by using the equations given in Fig. \ref{Fig:mosaic_astrometric_SNR}. For every contributing pointing for a given simulated source we randomly assign an RA and Dec astrometric offset given the S/N of the source, where again the S/N is the ratio of the injected integrated flux density to the local noise level in that particular pointing. We then shift the simulated source model  for that particular pointing by the offset. 

Having applied the smearing factors and astrometric errors to the individual pointing \textsc{clean} component or delta function models we combine the models to form the final mosaic model image of the field. Here we inject the weighted mean model image using the exact same weighting of the individual pointings as used to create the real mosaics at the position of the simulated source. The simulated sky model images are then convolved with the restoring beam of the mosaic (6$\arcsec$ or 9$\arcsec$) before being added to the \textsc{PyBDSF} residual mosaic image to produce the final simulated image.

The simulations were performed on 100 customised LoTSS-DR3 mosaics, each covering a circular area of approximately 13 square degrees and centred on the most sensitive individual LoTSS pointings. For each mosaic we performed ten runs to improve the statistics. To simplify the identification of injected sources in the analysis we only inject a fifth of the sources expected from T-RECS and ensure that all injected sources are separated by at least 15$\arcsec$.

To analyse the simulated images, we ran the same \textsc{PyBDSF} source finding procedure as used for the real mosaic images. However, we provide \textsc{PyBDSF} with the noise map from the real mosaic images to ensure the exact same noise properties are used for the cataloguing. Finally, for each injected source we assessed whether or not it is detected by seeing if it is within 3$\arcsec$ of the peak of the injected convolved model image of that source. If a source is detected we note that we are using the integrated flux density measurements for our analysis and these need altering because, depending on its detection significance, the source may have been injected with its (partially) smeared peak brightness (which in our simulations also reflects the reduction in the integrated flux density). We therefore optionally reverse the application of the smearing factors to obtain the true integrated flux density of the source which is what should be recovered in the mosaic image if \textsc{PyBDSF} is able to recover the full extent of the source despite the smearing effects, astrometric offsets and calibration errors. Thus our simulations mimic the impact on source finding from peak brightness reductions whilst optionally conserving the integrated flux density or allowing it to decrease proportionally to the reduction in peak brightness due to smearing and astrometric errors.

\begin{figure}
\includegraphics[width=\linewidth]{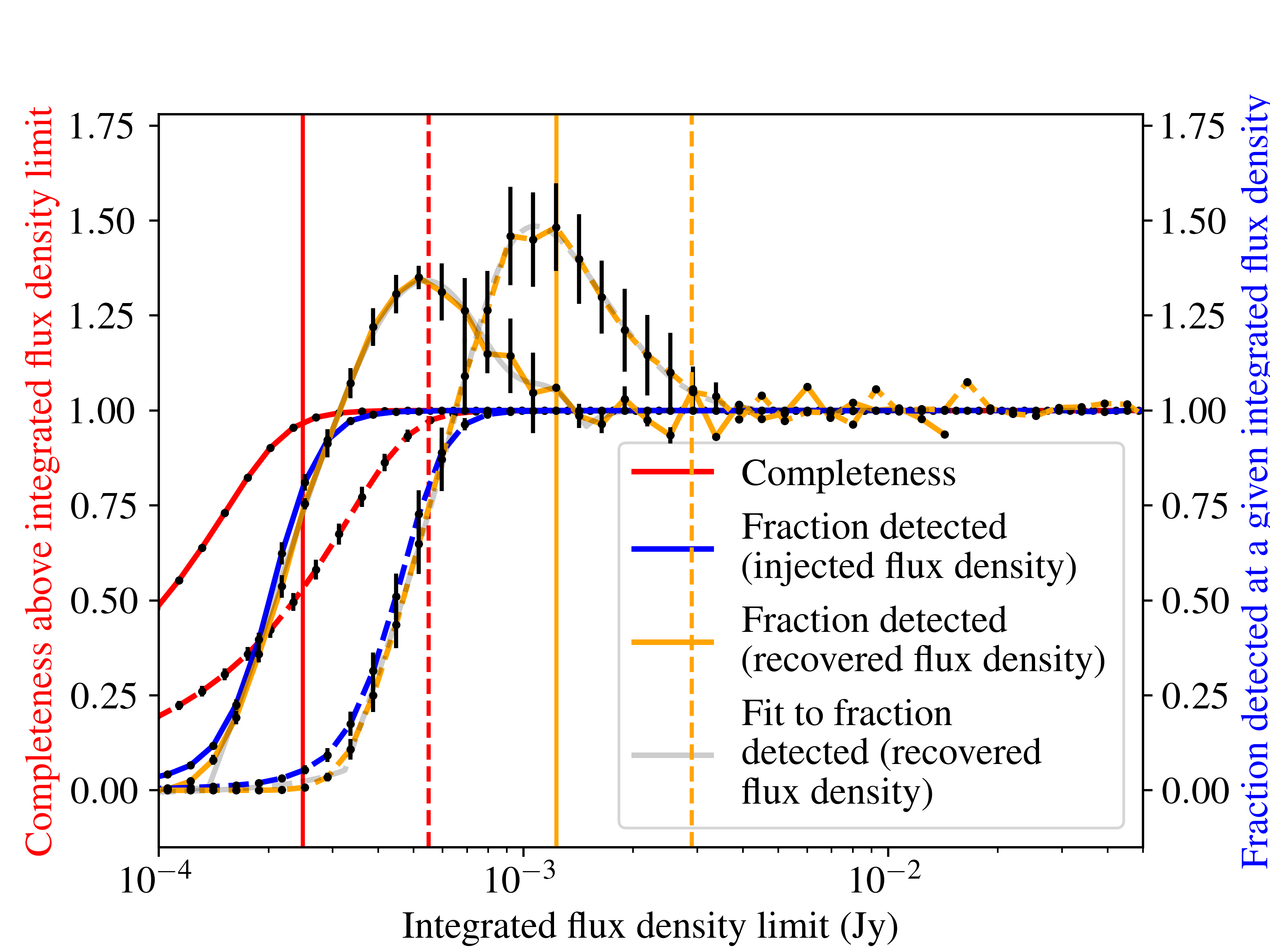}
\caption{Completeness as a function of integrated flux density and the fraction of sources detected at a given flux density, derived from mosaic simulations that include smearing and astrometric uncertainties. The results are based on 100 different simulated customised mosaic images (each with 10 runs), with error bars showing the standard deviation between the different mosaics simulated. In the simulations shown we have assumed \textsc{PyBDSF} is able to recover the full integrated flux density of sources in the presence of smearing and astrometric errors. Solid curves correspond to  the 14.3 deg$^2$ region with noise levels of 27-31$\mu$Jy beam$^{-1}$, while dashed curves represent the 85.0 deg$^2$ region with noise levels of 64-74$\mu$Jy beam$^{-1}$. The fraction of sources detected with a measured flux within given flux density range, relative to the number injected in that range, is modelled using a normalised sigmoid function modulated by a damped sine wave. The yellow vertical lines mark the highest integrated flux density where the model exceeds unity by more than 5\%, corresponding to 1.2\,mJy and 2.9\,mJy for the two noise bins and typically about 45 times the lower limit of the noise value for that bin. Above these thresholds the simulated source flux density distribution matches the recovered one. The red vertical lines indicate the integrated flux density where the completeness above the integrated flux density limit is 0.95, at  0.25\,mJy and 0.55\,mJy for these two noise bins -- typically 9 times the lower limit of the noise value for each bin.}
\label{Fig:completeness}
\end{figure}

The resulting \textsc{PyBDSF} catalogues of the simulated sources were used to derive the completeness. As the noise on a typical mosaic varies substantially across the image we measured the completeness at a given noise level rather than for a given mosaic image. In Fig. \ref{Fig:completeness}
we show the completeness, or the fraction of sources detected above a given injected integrated flux density as a function of the injected integrated flux density for all regions of the 100 simulated mosaics with \textsc{PyBDSF} noise levels in two given ranges, where here the noise level is adjusted from the raw \textsc{PyBDSF} noise images to also include smearing -- i.e. it reflects the noise level for smeared sources rather than unsmeared ones and so accounts for the fact that the noise down the primary beam is actually higher than the \textsc{PyBDSF} noise map value because of smearing. We find that the completeness exceeds 95\% at 9 times the RMS and that this behaviour is consistent with other noise bins. In the same figure we also show the fraction of sources detected at a given integrated flux density. Here we define the integrated flux density as either the integrated flux density of the injected source or the integrated flux density recovered. Firstly we see that we recover essentially all ($>$95\%) sources that are injected with an S/N above 11. However, for S/Ns 11 to 45 the number of sources recovered at a given integrated flux density is different from (higher than) the number of sources injected at that integrated flux density. This is because of \cite{Eddington_1913} bias and the steep increase in the number of sources with decreasing integrated flux density. This implies that the number of fainter sources detected due to them sitting on a positive noise spike far exceeds the number of brighter sources whose integrated flux densities are reduced if they sit on a negative noise spike, which results in a significant excess in the number of sources detected at moderate S/N (this effect is also seen in e.g. \citealt{Hale_2023}). For these two noise bins we find that the fraction of sources detected as a function of flux density does not settle at unity until an S/N of about 45. This behaviour is also seen for other noise bins. Hence both the completeness and the fraction of sources detected as a function of flux density and as a function of position on the sky can be obtained using the RMS noise levels across the sky.

\section{Image analysis}
\label{Sec:image_analysis}

\subsection{Euclidean-normalised differential source counts}
\label{Sec:source_counts}

We used the LoTSS-DR3 mosaics together with the LoTSS Deep ELAIS-N1 field (\citealt{Shimwell_2025}) to derive the Euclidean-normalised differential source counts. As is shown in Fig. \ref{Fig:completeness}, below an S/N of about 45, the number of sources injected into our mosaic images deviates from the number of sources recovered at the same flux density. Therefore, conservatively, we adopted an S/N threshold of 50 to derive the source counts directly from the full LoTSS-DR3 catalogue and the accompanying \textsc{PyBDSF} noise mosaics. To measure the source counts in a given integrated flux density bin, we first use the \textsc{PyBDSF} noise mosaics to calculate the total area in which a source of that integrated flux density would be detected with an S/N $\geq$ 50. The Euclidean-normalised source counts were then obtained by counting the number of sources within both that flux density bin and the corresponding sky area, and scaling accordingly. To estimate uncertainties, we divided the sky into 24 regions and computed the standard deviation of the different measurements in each integrated flux density bin. We also quantified the potential impact of $\pm$2\% systematic errors on the absolute flux-density scale of our images (\citealt{Hardcastle_2021,Shimwell_2022}). The Poisson errors on the source counts are generally negligible due to the large number of sources per bin, and a random $6\%$ error (as in Sec. \ref{Sec:mosaic_fluxscale}) on the LoTSS-DR3 integrated flux densities has minimal effect, as it largely cancels out statistically. However, given the survey sensitivity, the sky area to derive counts in this way rapidly declines below 5\,mJy. 

To probe the Euclidean-normalised differential source counts at lower integrated flux densities, completeness corrections are required. These corrections  account for the expected fraction of sources detected in a given flux density range relative to the number of sources that truly have flux densities within that range. Without such corrections, counts from different regions cannot be combined directly, as each region and flux density range requires a different scaling factor. We derive these scaling factors from the simulations described in Sec. \ref{Sec:completeness}, where they correspond to the correction needed to match the recovered source counts with the injected T-RECS counts. These correction factors are derived for each flux density bin of each of the 100 simulated customised mosaics. Example corrections are shown in Fig. \ref{Fig:completeness}, where the error bars reflect the pointing-to-pointing variations. 

\begin{figure}
\includegraphics[width=\linewidth]{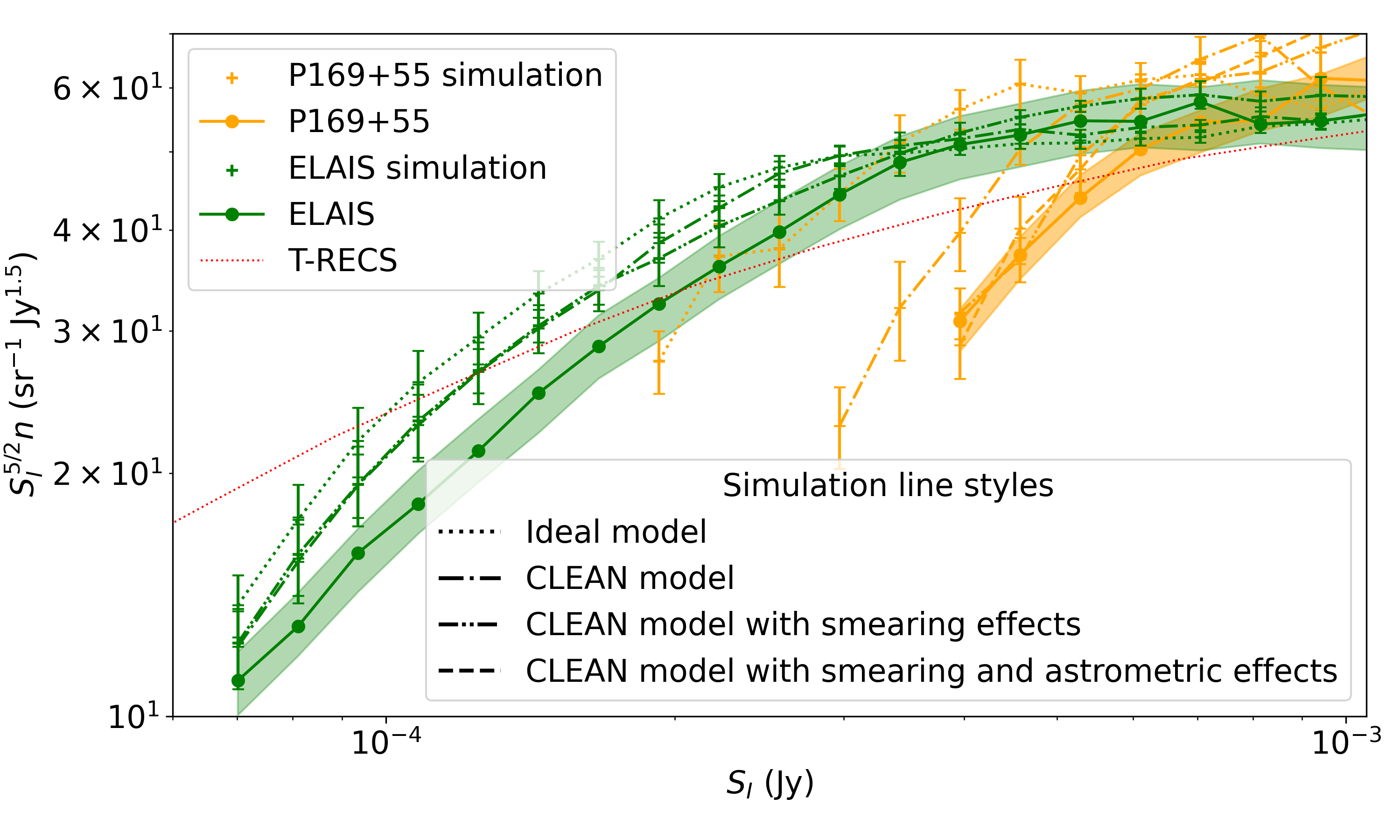}
\caption{Raw counts from a 13 square degree circular customised mosaic centred on pointing P169+55 (one of the deepest in LoTSS-DR3) and from the ELAIS-N1 LoTSS Deep field are shown as orange and green points, respectively. Shaded regions around the points indicate uncertainties due to $\pm$6\% and  $\pm$9\% flux density scale errors on the LoTSS-DR3 field and ELAIS-N1, respectively. The counts from completeness simulations (see Sec. \ref{Sec:completeness}) are also plotted, with the type of simulation indicated in the legend.  Error bars reflect $\pm1\sigma$ bootstrap uncertainties derived from the multiple runs of each simulation. Counts are shown down to the flux density where only 60\% of sources are recovered in the simulations and where the area (region with a RMS of at least 5 times lower than the flux density bin) used to derive the counts exceeds a square degree. The raw measured counts are also cut according to the same level using the simulations that include both smearing and astrometric effects. In these simulations we have assumed that \textsc{PyBDSF} is able to recover the full integrated flux density of sources in the presence of smearing and astrometric errors. The T-RECS source counts that are injected during the simulations are shown in red.}
\label{Fig:sourcecounts_simulations}
\end{figure}

To further illustrate the completeness corrections, example raw counts from two fields are shown together with source counts from our simulation in Fig. \ref{Fig:sourcecounts_simulations}. Here we show the results of several types of simulation to illustrate the impact of different effects; however, for deriving the source counts, we only use the simulations that include the \textsc{clean} components, smearing and astrometric effects, as these best match the observations and account for the known biases. Formal uncertainties in the simulations are estimated by bootstrap resampling over the different simulation runs of the same field. While these are displayed in Fig. \ref{Fig:sourcecounts_simulations}, the true uncertainties in the completeness corrections remain difficult to assess. For instance, the accuracy of the T-RECS source counts at these flux densities affects the derived completeness corrections due to \cite{Eddington_1913} bias. Furthermore, the simulations involve several approximations: the smearing could be modelled with a more realistic distortion of the source; the astrometric errors could be field specific; source blending or association uncertainties could be included; and ideally the simulations would be done by injecting appropriately distorted sources into the $uv$ data and reprocessing it (see e.g \citealt{Shimwell_2022}).

We applied the derived completeness corrections to the source counts from each noise-bin region of each mosaic, then combined them to form the final source counts. Noise-bin regions and flux density ranges where fewer than 60\% of sources were detected are excluded. Due to the challenges in quantifying uncertainties in the completeness correction,  we instead computed the standard deviation between the counts from different mosaics for each flux density bin. We also accounted for the uncertainty in the fraction of the total integrated flux density that \textsc{PyBDSF} recovers in the presence of smearing and astrometric errors, as well as indicating the limits that would correspond to $\pm$2\% systematic errors in the absolute flux density scale of our images.

Finally, we incorporated the counts from the ELAIS-N1 (\citealt{Shimwell_2025}) LoTSS Deep image using the same procedure as for the deep customised LoTSS-DR3 mosaics. The only differences are that astrometric errors are not included, since ELAIS-N1 is a single field calibrated against the same sky model. In addition,  we do not split the ELAIS-N1 image into multiple regions for the error analysis; instead, we adopt a uniform fractional error of 14\%, consistent with the typical error for the integrated flux density bins of the deep LoTSS-DR3 mosaics. We also indicate the limits that correspond to a $\pm$9\% systematic error on the absolute flux density scale of ELAIS-N1 (see \citealt{Shimwell_2025}).

\begin{figure}
\includegraphics[width=\linewidth]{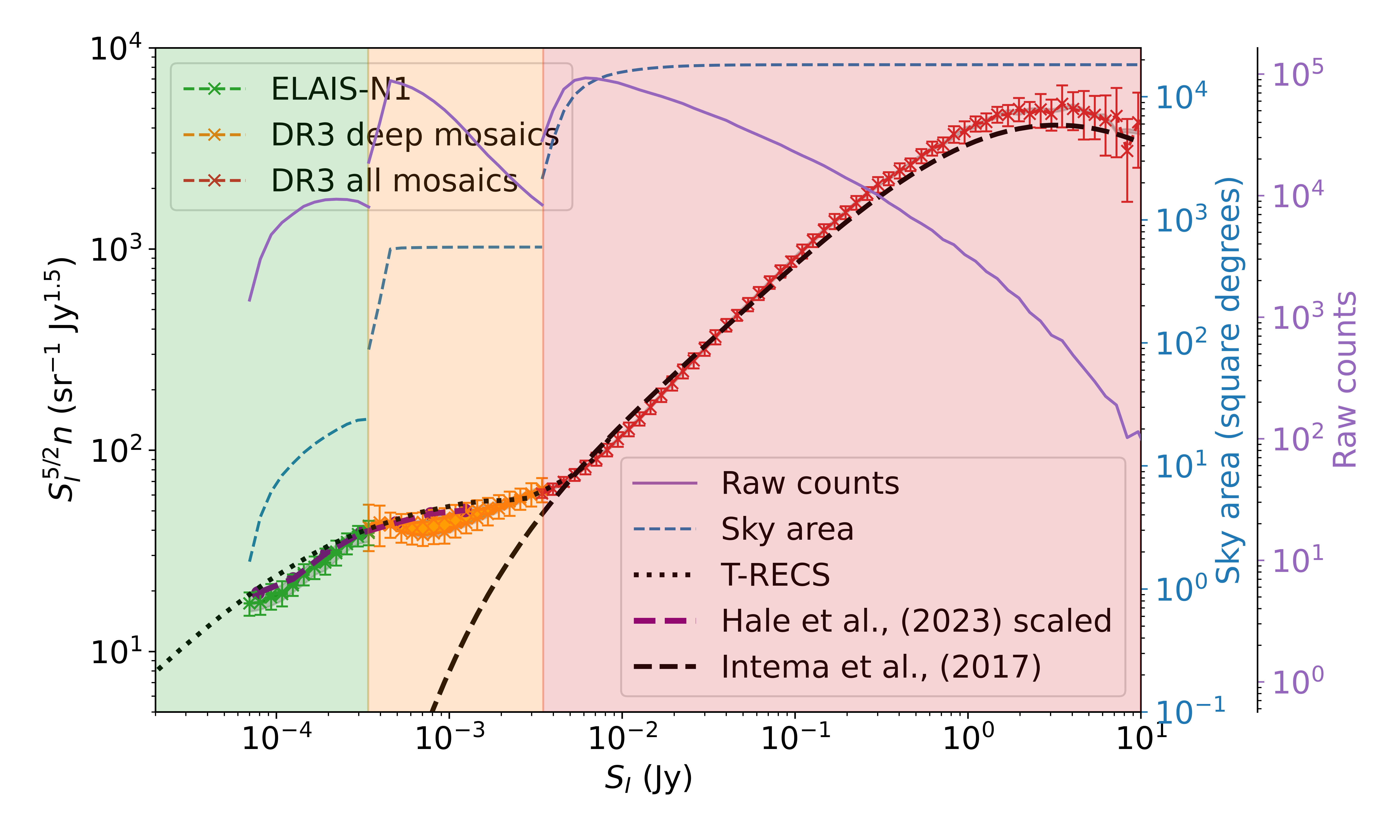}
\caption{Euclidean-normalised differential source counts from the LoTSS-DR3 and the ELAIS-N1 LoTSS Deep field. Green and orange markers correspond to counts derived from ELAIS-N1 and from the 100 customised LoTSS-DR3 mosaics around the deepest pointings with completeness simulations, respectively. These counts are corrected by the fraction of sources recovered in a particular flux density bin compared to the number injected in that bin during the simulations. Counts from a mosaic region and flux density bin are only included if the simulations indicate that at least 60\% of sources are recovered. Red markers show the uncorrected counts from all LoTSS-DR3 mosaics in regions where sources are detected at a significance exceeding 50. The error bars on the uncorrected LoTSS-DR3 counts are derived from splitting the northern sky into 24 regions and taking the standard deviation of the observed counts in a particular bin. The error bars on the 100 deepest LoTSS-DR3 mosaic counts are the standard deviation of the counts from the different mosaics in each bin. For ELAIS-N1 the error bars shown are the typical (14\%) error bar obtained from the variation in the LoTSS-DR3 deep mosaics over the sky. The filled orange and green regions illustrate the impact of \textsc{PyBDSF} on recovering the integrated flux density for faint, smeared, and partially deconvolved sources in our ELAIS-N1 and 100 customised LoTSS-DR3 mosaics around the deepest pointings, respectively. The lower boundary corresponds to full recovery, while the upper boundary represents only partial recovery, proportional to the reduction in peak brightness caused by smearing and astrometric errors. The shaded grey region around the measured source counts shows the systematic flux density scale errors. For LoTSS-DR3 this uncertainty is $\pm2\%$, which is difficult to discern at the scale of the figure, while for ELAIS-N1 it is $\pm9\%$. The dashed and dotted black lines show the \cite{Intema_2017} and T-RECS counts, respectively, where the \cite{Intema_2017} counts are severely impacted by incompleteness below $\sim$5\,mJy. The solid purple and dashed blue lines show the raw number of counts as well as the area used for the counts in a particular flux density bin. The \cite{Hale_2023} 1.4GHz counts for the COSMOS field with SIMBA completeness corrections and scaled by a spectral index of $-0.6$ are shown in dashed purple.}
\label{Fig:sourcecounts}
\end{figure}

A detailed interpretation of the LoTSS-DR3 source counts is left to a future publication. Overall, the combined Euclidean-normalised source counts from all LoTSS-DR3, the 100 customised LoTSS-DR3 mosaics around the deepest pointings, and ELAIS-N1 are shown in Fig. \ref{Fig:sourcecounts} and Tab. \ref{tab:source_counts}. The source counts probe integrated flux densities above 70\,$\mu$Jy/beam and are a good match to the wide-area counts from \cite{Intema_2017} at high integrated flux densities, although showing slight differences in detail comparable to those that have been seen in previous wide-area source count analysis \citep{Hardcastle_2021}. At low integrated flux densities, they also match the T-RECS simulations reasonably well. They can also be compared to deep counts at higher frequencies, such as the 1.4\,GHz counts from \cite{Hale_2023}. Naively, assuming a single spectral index $\alpha$ for all radio sources, we fit $\alpha$ to align the faint bins of the 1.4\,GHz COSMOS field counts corrected for completeness using SIMBA simulations (\citealt{Dave_2019}) with our ELAIS-N1 counts, finding $\alpha=-0.6$. This spectral index is flatter than that expected from typical sources between these two frequency ranges as seen in wide area surveys (e.g. \citealt{Bohme_2023}), but at low flux densities, star-forming galaxies begin to dominate the population and tend to have flatter spectral indexes (e.g. \citealt{Williams_2021} and \citealt{An_2024}). The agreement between LoTSS-DR3, ELAIS LoTSS Deep and existing studies supports the completeness assessment conducted in Sec. \ref{Sec:completeness} and the overall fidelity of our images.

\section{Future prospects}
\label{Sec:future_prospects}

LOFAR is currently undergoing an upgrade to LOFAR2.0 and once this is completed in 2026 we aim to resume LoTSS observations to achieve full coverage of the northern hemisphere. LOFAR2.0 will provide 192\,MHz of simultaneous bandwidth, enabling the use of four simultaneous 48\,MHz bandwidth beams instead of the current two. Completing the remaining 534 survey pointings will thus require 1,230\,hrs of LOFAR2.0 observing.

In this publication we have focussed on the extragalactic Stokes I continuum images at $6\arcsec$ -- $9\arcsec$ resolution which are the primary LoTSS data products. However, we have also produced and made publicly available 20$\arcsec$ resolution Stokes I images, as well as Stokes Q and U cubes and Stokes V images, for probing extragalactic and Galactic linearly or circularly polarised sources. We are also in the process of generating snapshot Stokes I and Stokes V images (with 8\,s and 2\,min integrations) spanning the entire 12,950 hrs of LoTSS-DR3 data. Very low-resolution (60$\arcsec$) images of the sky, with and without detected sources removed, are being created to probe faint diffuse Galactic and extragalactic structures. Furthermore, we are extracting dynamic spectra for hundreds of thousands of targeted sources in the dataset. We are also embarking on substantial projects to refine the calibration towards especially interesting targets through the extraction and self-calibration procedure, to produce higher-resolution (6$\arcsec$) Stokes Q and U cubes and to more carefully characterise the complex Galactic emission. Each of these aspects will be the subject of future publications. We note, however, that for LoTSS-DR2 -- which was created using the same data processing pipeline -- we conducted a detailed characterisation of the polarisation product quality (\citealt{Callingham_2023}; \citealt{Erceg_2022}; \citealt{Shimwell_2022}; \citealt{OSullivan_2023}), the snapshot images (\citealt{deRuiter_2024}), the very low resolution maps (\citealt{Oei_2022}), and the refined calibration procedure (\citealt{vanWeeren_2021}); these studies already provide valuable insights into the data quality to be expected from DR3. The post processing can all be done from the publicly available LoTSS-DR3 datasets using routines in the DDF-pipeline repository, because we preserve the visibilities, associated calibration solutions and necessary image products.

Wide-area optical identifications for LoTSS-DR3 do not yet exist,
other than those already carried out for DR2 \citep{Hardcastle_2023}.
Because DR3 covers so much of the Northern sky, it includes areas
where the best optical coverage at present is the comparatively
shallow Pan-STARRS data, as well as regions of the Galactic plane
where optical identifications will be challenging due to dust
obscuration. It is likely that there will never be a complete optical
identification of the DR3 sources, but expect to produce a catalogue of optical
identifications at high Galactic latitude where good optical data are
available. Our immediate goal will be to match to the deep space- and
ground-based observations to be made available over many thousands of
square degrees by the \textit{Euclid} mission and its programme of
ground-based supporting observations.

A longer-term aim is to systematically exploit the international LOFAR stations, which requires using the full time and frequency resolution of the LoTSS data (typically 1s and 12.1875\,kHz) to avoid otherwise prohibitive time- and frequency-smearing effects. The international stations have already been used to create high angular resolution ($\sim0.3\arcsec$) of individual targets  (e.g. \citealt{Morabito_2022}, \citealt{Timmerman_2022}, \citealt{vanWeeren_2024} and \citealt{De_Rubeis_2025}) or even entire pointings (e.g. \citealt{Sweijen_2022},  \citealt{deJong_2024}, Bondi et al. in prep. and \citealt{Escott_submitted}), but no wide-area images spanning hundreds of square degrees have yet been produced at this angular resolution. Initial efforts are underway, including a survey producing postage-stamp images of selected sources across tens of square degrees, and calibrators have already been identified across a large fraction of the northern sky \citep{Jackson_2016,Jackson_2022}. In addition, the efficiency, automation and image quality are continuing to improve (e.g. \citealt{deJong_2025}). However, a limitation arises from the angular separation of LoTSS pointings (2.58$^\circ$): time- and frequency-smearing effects become increasingly severe away from the pointing centre, and the international LOFAR stations have a smaller field of view (FWHM 2.07$^\circ$ at 150\,MHz, compared to 3.8$^\circ$ for the remote and core stations in HBA Dual Inner mode which was used for LoTSS observations). Together, these factors lead to large sensitivity variations in wide-area mosaics that combine many pointings. To address this, we are preparing the International LOFAR Two-metre Sky Survey (ILoTSS), a LOFAR2.0 project designed to achieve more uniform wide-area sensitivity at the full angular resolution offered by the international stations. ILoTSS will reuse the existing LoTSS dataset while adding a new pointing halfway between each original LoTSS pointing (providing four times as many pointings per unit sky area). This tighter grid will mitigate both smearing effects and field-of-view limitations, enabling close to uniform sensitivity coverage over very wide areas with a typical RMS noise of 30 $\mu$Jy beam$^{-1}$ at the full resolution of 0.3 arcsec. ILoTSS will thus provide a uniquely good view of the northern extragalactic sky at low frequencies. Observations for the project are expected to start in 2027.

\section{Summary}
\label{Sec:Summary}

We have described the public release of LoTSS-DR3, which consists of images, catalogues, and data products that span 88\% of the northern sky. This is the largest radio wavelength survey ever carried out in terms of total number of sources detected (13,667,877) and almost four times larger than our previous data release. The mosaicked 120-168\,MHz images have a median sensitivity of 92\,$\mu$Jy beam$^{-1}$ at an angular resolution of 6$\arcsec$ (9$\arcsec$ below declination 10$^\circ$). Analysis of the images demonstrates an astrometric accuracy that follows the expected noise-like behaviour but with an additional systematic offset of at least 0.24$\arcsec$ due to calibration inaccuracies. The random flux density scale error is 6\%, while the systematic offset was previously shown to be within 2\% (\citealt{Hardcastle_2021}, \citealt{Shimwell_2022}). To achieve this level of accuracy, our astrometry has been matched to Pan-STARRS, the overall flux density scale has been tied to NVSS, and flux density scale variations across the field of view have been reduced by aligning neighbouring LoTSS pointings. A completeness analysis accounting for smearing effects, astrometric errors, and realistic source models indicates that we detect approximately 95\% of radio sources with flux densities exceeding 9 times the local RMS. Using these completeness simulations, we have derived the Euclidean-normalised differential source counts over five orders of magnitude, finding good agreement with previous studies and simulations. 

\section*{Data availability}
\label{Sec:release_details}

The LoTSS-DR3 data products are publicly available and can be accessed via the LOFAR Surveys webpage (\url{https://lofar-surveys.org/dr3_release.html}), the SURF data repository (\url{https://doi.org/10.25606/SURF.lotss-dr3}), or the ASTRON Virtual Observatory (\url{https://vo.astron.nl}). The source catalogue is also hosted at the Strasbourg Astronomical Data Center (CDS) and available via ftp
to \url{cdsarc.u-strasbg.fr (130.79.128.5)} or via \url{http://cdsweb.u-strasbg.fr/cgi-bin/qcat?J/A+A/}. The data collection in this release has the Digital Object Identifier (DOI) 10.25606/SURF.LoTSS-DR3, and each individual data product is assigned a unique ePIC persistent identifier (ePIC-PID). The data products include:
\begin{itemize}
\item 2,551 individual field full depth Stokes I 6$\arcsec$ and 20$\arcsec$ resolution continuum images (9$\arcsec$ and 20$\arcsec$ for declinations below 10$^\circ$) each covering $8^{\circ}\times8^{\circ}$  and with a median sensitivity (excluding fields within 10$^\circ$ of the galactic plane) of 112\,$\mu$Jy/beam in the higher resolution maps.
\item 2,551 sets of individual field Stokes V 6$\arcsec$ (9$\arcsec$ for declinations below 10$^\circ$) resolution continuum images where an image is provided for each epoch (up to 6 epochs per field).
\item 2,551 individual field Stokes Q and U 20$\arcsec$ and 4$\arcmin$ resolution cubes (typically 462 channels of width 97.6\,kHz).
\item 2,551 sets of individual field $uv$ data and direction-dependent calibration solutions (at a time resolution of 8\,s and frequency resolution of 97.6\,kHz).
\item 1,580 mosaicked Stokes I 6$\arcsec$ and 20$\arcsec$ resolution
  continuum images centred on HEALPix pixels with {\tt NSIDE=16} (9$\arcsec$ and 20$\arcsec$ for declinations below 10$^\circ$).
  \item A merged \textsc{PyBDSF} catalogue of 13,667,877 sources and the corresponding 16,943,656 Gaussian components derived from all Stokes I 6$\arcsec$ resolution mosaics. The columns are described in Tab. \ref{tab:catalogue_columns} and are a subset of those provided by default by \textsc{PyBDSF}, including position, integrated flux density, peak brightness, size, and estimated statistical errors on all parameters (but not including flux density scale or astrometric alignment errors, which should be added appropriately during analysis). We add additional columns with the identity of the mosaic in which a source was detected and the angular resolution of the mosaic image.
\end{itemize}

The final DDF-pipeline data products for the 2,551 fields processed
are 590\,TB in size, of which 8\% is Stokes I images, 24\% is Stokes
Q, U, and V images and 67\% is $uv$ data and associated calibration
solutions. The 18.6\,PB of uncalibrated LoTSS data that were used for
LoTSS-DR3 are also publicly available and can be retrieved from the
LOFAR LTA.

\label{lastpage}

\onecolumn
\begin{appendix}
\section{Catalogue column descriptions}
\label{appendix_pybdsf}

\begin{table}[htbp]
\centering
\caption{Description of the columns in the merged mosaic source catalogue (S) and the Gaussian-component catalogue (G).}
\label{tab:catalogue_columns}
\begin{tabular}{lll p{9cm}}
\hline
Column name & Units & Catalogue(s) & Description \\
\hline
Source\_Name & -- & S, G & The radio name of the source in the format "ILTJHHMMSS.s+DDMMSS" \\
RA & deg & S, G & Right ascension (J2000) \\
E\_RA & arcsec & S, G & Uncertainty in right ascension \\
DEC & deg & S, G & Declination (J2000) \\
E\_DEC & arcsec & S, G & Uncertainty in declination \\
Total\_flux & mJy & S, G & Integrated Stokes I flux density at 144 MHz \\
E\_Total\_flux & mJy & S, G & Uncertainty in integrated Stokes I flux density \\
Peak\_flux & mJy beam$^{-1}$ & S, G & Peak Stokes I flux density at 144 MHz per beam of the source \\
E\_Peak\_flux & mJy beam$^{-1}$ & S, G & Uncertainty in peak Stokes I flux density per beam of the source \\
Maj & arcsec & S, G & Fitted FWHM of the major axis of the source \\
E\_Maj & arcsec & S, G & Uncertainty in FWHM of the fitted major axis \\
Min & arcsec & S, G & Fitted FWHM of the minor axis of the source \\
E\_Min & arcsec & S, G & Uncertainty in FWHM of the fitted minor axis \\
PA & deg & S, G & Fitted position angle of the major axis of the source measured east of north \\
E\_PA & deg & S, G & Uncertainty in fitted position angle of the major axis of the source \\
DC\_Maj & arcsec & S, G & Deconvolved FWHM of the major axis of the source \\
E\_DC\_Maj & arcsec & S, G & Uncertainty in deconvolved FWHM of the major axis \\
DC\_Min & arcsec & S, G & Deconvolved FWHM of the minor axis of the source \\
E\_DC\_Min & arcsec & S, G & Uncertainty in deconvolved FWHM of the minor axis \\
DC\_PA & deg & S, G & Deconvolved position angle of the major axis of the source measured east of north \\
E\_DC\_PA & deg & S, G & Uncertainty in deconvolved position angle of the major axis of the source \\
Isl\_Total\_flux & mJy & S, G & Integrated Stokes I flux density of the island in which the source is located \\
E\_Isl\_Total\_flux & mJy & S, G & Uncertainty in integrated Stokes I flux density of the island \\
Isl\_rms & mJy beam$^{-1}$ & S, G & Average background RMS value of the island in which the source is located \\
S\_Code & -- & S, G & Source classification code. S = a single-Gaussian source that is the only source in the island; C = a single-Gaussian source in an island with other sources; M = a multi-Gaussian source \\
HEALPIX & -- & S, G & HEALPix pixel index of the source position, computed at $\mathrm{NSIDE}=16$. \\
Resolution & arcsec & S, G & The angular resolution of the mosaic image from which the source was extracted \\
Gaus\_id & -- & G & Unique identifier for Gaussian component, starting from zero \\
Wave\_id & -- & G & Wavelet scale from which the source was extracted, starting from zero \\
Xposn & pix & G & X image coordinate of the source \\
E\_Xposn & pix & G & Uncertainty in X image coordinate \\
Yposn & pix & G & Y image coordinate of the source \\
E\_Yposn & pix & G & Uncertainty in Y image coordinate \\
Wave\_Isl\_rms & Jy beam$^{-1}$ & G & Average background RMS value of the island in the wavelet image in which the Gaussian was fit \\
Wave\_Isl\_mean & Jy beam$^{-1}$ & G & Average background mean value of the island in the wavelet image in which the Gaussian was fit \\
\hline
\end{tabular}
\tablefoot{The listed columns are a subset of those output by \textsc{PyBDSF}, with additional columns identifying the mosaic in which each source was detected and providing the mosaic angular resolution. Quoted uncertainties are the statistical errors reported by \textsc{PyBDSF}; they do not include contributions from the uncertainties in the flux-density scale or astrometric alignment, which should be accounted for during analysis. The catalogues contain 13,667,877 sources corresponding to 16,943,656 Gaussian components. }
\end{table}

\newpage

\section{Euclidean-normalised differential source counts table}

\setlength{\LTcapwidth}{\textwidth}
\begin{ThreePartTable}
\begin{TableNotes}[flushleft]
\footnotesize
\item \textbf{Notes.} Here $S_I$ is the mid-point integrated flux density of each bin, and the $S_I$ range gives the flux-density limits of the bin. The Type column indicates the image used for the measurements, where $\rm{E-N1}$, $\rm{M_C}$, and $\rm{M_R}$ correspond to the ELAIS-N1 image, deep LoTSS-DR3 mosaics with completeness corrections, and all LoTSS-DR3 mosaics without any completeness correction, respectively. $N$ is the number of sources detected in a given flux-density bin, Area is the sky area used for the measurement, $F$ is the fraction of sources recovered in a particular flux density bin compared to the number injected in that bin during the simulations, and Counts gives the Euclidean-normalised 144\,MHz differential source count after correcting for $F$. The errors $E_g$, $E_s$, and $E_p$ represent the field-to-field variations (or 14\% for ELAIS-N1), the effect of systematic flux-density scale uncertainties (9\% for ELAIS-N1 and 2\% for LoTSS-DR3), and the Poisson uncertainty on the number of detected sources, respectively.
\end{TableNotes}
\begin{longtable}{
  d{3.3}       
  c            
  c            
  d{3.5}       
  d{3.3}       
  d{1.1}       
  d{3.2}       
  d{2.1}       
  d{2.2}       
  d{1.2}       
}
\caption{144\,MHz Euclidean-normalised differential source counts derived from LoTSS-DR3 and the LoTSS Deep ELAIS-N1 field.}
\label{tab:source_counts}\\

\toprule
\multicolumn{1}{c}{$S_I$} &
\multicolumn{1}{c}{$S_I$ range} &
\multicolumn{1}{c}{Type} &
\multicolumn{1}{c}{$N$} &
\multicolumn{1}{c}{Area} &
\multicolumn{1}{c}{$F$} &
\multicolumn{1}{c}{Counts} &
\multicolumn{1}{c}{$E_g$} &
\multicolumn{1}{c}{$E_s$} &
\multicolumn{1}{c}{$E_p$} \\
\multicolumn{1}{c}{[mJy]} &
\multicolumn{1}{c}{[mJy]} &
\multicolumn{1}{c}{} &
\multicolumn{1}{c}{[\#]} &
\multicolumn{1}{c}{[\si{\square\degree}]} &
\multicolumn{1}{c}{} &
\multicolumn{1}{c}{\shortstack{sr$^{-1}$ \\ Jy$^{1.5}$}} &
\multicolumn{1}{c}{\shortstack{sr$^{-1}$ \\ Jy$^{1.5}$}} &
\multicolumn{1}{c}{\shortstack{sr$^{-1}$ \\ Jy$^{1.5}$}} &
\multicolumn{1}{c}{[\shortstack{sr$^{-1}$ \\ Jy$^{1.5}$}]} \\
\midrule
\endfirsthead

\toprule
\multicolumn{10}{c}{\tablename~\thetable{} -- continued from previous page} \\
\midrule
\multicolumn{1}{c}{$S_I$} &
\multicolumn{1}{c}{$S_I$ range} &
\multicolumn{1}{c}{Type} &
\multicolumn{1}{c}{$N$} &
\multicolumn{1}{c}{Area} &
\multicolumn{1}{c}{$F$} &
\multicolumn{1}{c}{Counts} &
\multicolumn{1}{c}{$E_g$} &
\multicolumn{1}{c}{$E_s$} &
\multicolumn{1}{c}{$E_p$} \\
\multicolumn{1}{c}{[mJy]} &
\multicolumn{1}{c}{[mJy]} &
\multicolumn{1}{c}{} &
\multicolumn{1}{c}{[\#]} &
\multicolumn{1}{c}{[\si{\square\degree}]} &
\multicolumn{1}{c}{} &
\multicolumn{1}{c}{\shortstack{sr$^{-1}$ \\ Jy$^{1.5}$}} &
\multicolumn{1}{c}{\shortstack{sr$^{-1}$ \\ Jy$^{1.5}$}} &
\multicolumn{1}{c}{\shortstack{sr$^{-1}$ \\ Jy$^{1.5}$}} &
\multicolumn{1}{c}{[\shortstack{sr$^{-1}$ \\ Jy$^{1.5}$}]} \\
\midrule
\endhead

\midrule
\multicolumn{10}{r}{Continued on next page}
\endfoot

\bottomrule
\insertTableNotes
\endlastfoot
0.07 & \numrange{0.065}{0.075} & $\rm{E-N1}$ & $1.38 \times 10^{3}$ & 1.66 & 0.64 & 17 & 2.3 & 1.6 & 0.37 \\
0.081 & \numrange{0.075}{0.087} & $\rm{E-N1}$ & $3.00 \times 10^{3}$ & 3.85 & 0.74 & 18 & 2.3 & 1.6 & 0.27 \\
0.094 & \numrange{0.087}{0.1} & $\rm{E-N1}$ & $4.77 \times 10^{3}$ & 6.16 & 0.83 & 19 & 2.5 & 1.7 & 0.25 \\
0.11 & \numrange{0.1}{0.12} & $\rm{E-N1}$ & $6.00 \times 10^{3}$ & 8.37 & 0.93 & 20 & 2.6 & 1.8 & 0.25 \\
0.12 & \numrange{0.12}{0.13} & $\rm{E-N1}$ & $7.03 \times 10^{3}$ & 10.5 & 1 & 21 & 2.8 & 1.9 & 0.25 \\
0.14 & \numrange{0.13}{0.16} & $\rm{E-N1}$ & $8.15 \times 10^{3}$ & 12.8 & 1 & 24 & 3.2 & 2.1 & 0.27 \\
0.17 & \numrange{0.16}{0.18} & $\rm{E-N1}$ & $8.82 \times 10^{3}$ & 15.0 & 1.1 & 26 & 3.4 & 2.4 & 0.29 \\
0.19 & \numrange{0.18}{0.21} & $\rm{E-N1}$ & $9.22 \times 10^{3}$ & 17.3 & 1.1 & 29 & 3.8 & 2.5 & 0.32 \\
0.22 & \numrange{0.21}{0.24} & $\rm{E-N1}$ & $9.33 \times 10^{3}$ & 19.5 & 1.2 & 31 & 4.1 & 2.8 & 0.35 \\
0.26 & \numrange{0.24}{0.28} & $\rm{E-N1}$ & $9.26 \times 10^{3}$ & 21.8 & 1.2 & 34 & 4.5 & 3.1 & 0.38 \\
0.3 & \numrange{0.28}{0.32} & $\rm{E-N1}$ & $8.94 \times 10^{3}$ & 23.5 & 1.2 & 37 & 4.9 & 3.4 & 0.43 \\
\midrule
0.34 & \numrange{0.32}{0.37} & $\rm{M_C}$ & $1.88 \times 10^{4}$ & 94.1 & 0.72--0.74 & 41--42 & 11 & 0.62 & 0.25 \\
0.4 & \numrange{0.37}{0.43} & $\rm{M_C}$ & $3.91 \times 10^{4}$ & 252 & 0.72--0.77 & 42--44 & 9.1 & 0.55 & 0.18 \\
0.46 & \numrange{0.43}{0.49} & $\rm{M_C}$ & $8.80 \times 10^{4}$ & 578 & 0.78--0.8 & 42--44 & 5.5 & 0.66 & 0.13 \\
0.53 & \numrange{0.49}{0.57} & $\rm{M_C}$ & $8.32 \times 10^{4}$ & 590 & 0.91--0.99 & 40--43 & 5 & 0.72 & 0.14 \\
0.61 & \numrange{0.57}{0.66} & $\rm{M_C}$ & $7.69 \times 10^{4}$ & 593 & 1--1.2 & 38--43 & 4.8 & 0.85 & 0.15 \\
0.71 & \numrange{0.66}{0.76} & $\rm{M_C}$ & $6.87 \times 10^{4}$ & 595 & 1.1--1.3 & 38--44 & 4.8 & 1.1 & 0.17 \\
0.81 & \numrange{0.76}{0.88} & $\rm{M_C}$ & $5.97 \times 10^{4}$ & 597 & 1.2--1.4 & 39--45 & 4.7 & 1.2 & 0.19 \\
0.94 & \numrange{0.88}{1} & $\rm{M_C}$ & $5.06 \times 10^{4}$ & 598 & 1.2--1.4 & 39--46 & 5.1 & 1.5 & 0.22 \\
1.1 & \numrange{1}{1.2} & $\rm{M_C}$ & $4.17 \times 10^{4}$ & 598 & 1.2--1.4 & 42--48 & 5 & 1.6 & 0.25 \\
1.3 & \numrange{1.2}{1.4} & $\rm{M_C}$ & $3.37 \times 10^{4}$ & 599 & 1.2--1.3 & 43--49 & 5.1 & 1.8 & 0.28 \\
1.5 & \numrange{1.4}{1.6} & $\rm{M_C}$ & $2.67 \times 10^{4}$ & 599 & 1.1--1.2 & 46--50 & 5.8 & 1.9 & 0.32 \\
1.7 & \numrange{1.6}{1.8} & $\rm{M_C}$ & $2.14 \times 10^{4}$ & 599 & 1.1--1.2 & 48--52 & 6 & 1.9 & 0.36 \\
1.9 & \numrange{1.8}{2.1} & $\rm{M_C}$ & $1.76 \times 10^{4}$ & 599 & 1.1 & 51--54 & 5.7 & 1.8 & 0.41 \\
2.2 & \numrange{2.1}{2.4} & $\rm{M_C}$ & $1.43 \times 10^{4}$ & 600 & 1--1.1 & 54--56 & 6.6 & 2.0 & 0.47 \\
2.6 & \numrange{2.4}{2.8} & $\rm{M_C}$ & $1.18 \times 10^{4}$ & 600 & 1 & 57--58 & 6.2 & 2.0 & 0.53 \\
3 & \numrange{2.8}{3.2} & $\rm{M_C}$ & $9.84 \times 10^{3}$ & 600 & 1 & 60--61 & 7.5 & 2.2 & 0.61 \\
\midrule
3.4 & \numrange{3.2}{3.7} & $\rm{M_R}$ & $2.84 \times 10^{4}$ & $2.14 \times 10^{3}$ & 1 & 61 & 3 & 1.2 & 0.36 \\
4 & \numrange{3.7}{4.3} & $\rm{M_R}$ & $4.99 \times 10^{4}$ & $4.45 \times 10^{3}$ & 1 & 64 & 4.2 & 1.3 & 0.29 \\
4.6 & \numrange{4.3}{4.9} & $\rm{M_R}$ & $7.49 \times 10^{4}$ & $7.63 \times 10^{3}$ & 1 & 70 & 3.9 & 1.4 & 0.25 \\
5.3 & \numrange{4.9}{5.7} & $\rm{M_R}$ & $8.86 \times 10^{4}$ & $1.04 \times 10^{4}$ & 1 & 75 & 5.1 & 1.5 & 0.25 \\
6.1 & \numrange{5.7}{6.6} & $\rm{M_R}$ & $9.28 \times 10^{4}$ & $1.23 \times 10^{4}$ & 1 & 82 & 6.4 & 1.6 & 0.27 \\
7.1 & \numrange{6.6}{7.6} & $\rm{M_R}$ & $9.16 \times 10^{4}$ & $1.38 \times 10^{4}$ & 1 & 90 & 6.7 & 1.8 & 0.3 \\
8.2 & \numrange{7.6}{8.8} & $\rm{M_R}$ & $8.83 \times 10^{4}$ & $1.49 \times 10^{4}$ & 1 & $1 \times 10^{2}$ & 7.3 & 2.0 & 0.34 \\
9.5 & \numrange{8.8}{10} & $\rm{M_R}$ & $8.47 \times 10^{4}$ & $1.56 \times 10^{4}$ & 1 & $1.1 \times 10^{2}$ & 9.6 & 2.3 & 0.39 \\
11 & \numrange{10}{12} & $\rm{M_R}$ & $7.93 \times 10^{4}$ & $1.62 \times 10^{4}$ & 1 & $1.3 \times 10^{2}$ & 10 & 2.5 & 0.45 \\
13 & \numrange{12}{14} & $\rm{M_R}$ & $7.40 \times 10^{4}$ & $1.67 \times 10^{4}$ & 1 & $1.40 \times 10^{2}$ & 11 & 2.9 & 0.53 \\
15 & \numrange{14}{16} & $\rm{M_R}$ & $6.96 \times 10^{4}$ & $1.7 \times 10^{4}$ & 1 & $1.6 \times 10^{2}$ & 13 & 3.3 & 0.62 \\
17 & \numrange{16}{18} & $\rm{M_R}$ & $6.55 \times 10^{4}$ & $1.73 \times 10^{4}$ & 1 & $1.9 \times 10^{2}$ & 14 & 3.8 & 0.73 \\
19 & \numrange{18}{21} & $\rm{M_R}$ & $6.12 \times 10^{4}$ & $1.76 \times 10^{4}$ & 1 & $2.2 \times 10^{2}$ & 18 & 4.3 & 0.87 \\
23 & \numrange{21}{24} & $\rm{M_R}$ & $5.71 \times 10^{4}$ & $1.77 \times 10^{4}$ & 1 & $2.5 \times 10^{2}$ & 20 & 4.9 & 1 \\
26 & \numrange{24}{28} & $\rm{M_R}$ & $5.23 \times 10^{4}$ & $1.79 \times 10^{4}$ & 1 & $2.8 \times 10^{2}$ & 24 & 5.6 & 1.2 \\
30 & \numrange{28}{32} & $\rm{M_R}$ & $4.84 \times 10^{4}$ & $1.80 \times 10^{4}$ & 1 & $3.2 \times 10^{2}$ & 24 & 6.4 & 1.4 \\
35 & \numrange{32}{37} & $\rm{M_R}$ & $4.48 \times 10^{4}$ & $1.80 \times 10^{4}$ & 1 & $3.6 \times 10^{2}$ & 29 & 7.3 & 1.7 \\
40 & \numrange{37}{43} & $\rm{M_R}$ & $4.15 \times 10^{4}$ & $1.81 \times 10^{4}$ & 1 & $4.2 \times 10^{2}$ & 30 & 8.3 & 2.1 \\
46 & \numrange{43}{50} & $\rm{M_R}$ & $3.75 \times 10^{4}$ & $1.81 \times 10^{4}$ & 1 & $4.7 \times 10^{2}$ & 29 & 9.4 & 2.4 \\
53 & \numrange{50}{57} & $\rm{M_R}$ & $3.42 \times 10^{4}$ & $1.81 \times 10^{4}$ & 1 & $5.3 \times 10^{2}$ & 40 & 11 & 2.9 \\
62 & \numrange{57}{66} & $\rm{M_R}$ & $3.13 \times 10^{4}$ & $1.82 \times 10^{4}$ & 1 & $6.0 \times 10^{2}$ & 44 & 12 & 3.4 \\
71 & \numrange{66}{77} & $\rm{M_R}$ & $2.86 \times 10^{4}$ & $1.82 \times 10^{4}$ & 1 & $6.8 \times 10^{2}$ & 49 & 14 & 4 \\
82 & \numrange{77}{89} & $\rm{M_R}$ & $2.62 \times 10^{4}$ & $1.82 \times 10^{4}$ & 1 & $7.7 \times 10^{2}$ & 55 & 15 & 4.8 \\
95 & \numrange{89}{1e+02} & $\rm{M_R}$ & $2.35 \times 10^{4}$ & $1.82 \times 10^{4}$ & 1 & $8.7 \times 10^{2}$ & 52 & 17 & 5.6 \\
$1.1 \times 10^{2}$ & \numrange{1e+02}{1.2e+02} & $\rm{M_R}$ & $2.14 \times 10^{4}$ & $1.82 \times 10^{4}$ & 1 & $9.7 \times 10^{2}$ & 77 & 20 & 6.7 \\
$1.3 \times 10^{2}$ & \numrange{1.2e+02}{1.4e+02} & $\rm{M_R}$ & $1.95 \times 10^{4}$ & $1.82 \times 10^{4}$ & 1 & $1.1 \times 10^{3}$ & 80 & 22 & 7.9 \\
$1.5 \times 10^{2}$ & \numrange{1.4e+02}{1.6e+02} & $\rm{M_R}$ & $1.76 \times 10^{4}$ & $1.82 \times 10^{4}$ & 1 & $1.2 \times 10^{3}$ & 89 & 25 & 9.3 \\
$1.7 \times 10^{2}$ & \numrange{1.6e+02}{1.8e+02} & $\rm{M_R}$ & $1.57 \times 10^{4}$ & $1.82 \times 10^{4}$ & 1 & $1.4 \times 10^{3}$ & $1.2 \times 10^{2}$ & 27 & 11 \\
$2 \times 10^{2}$ & \numrange{1.8e+02}{2.1e+02} & $\rm{M_R}$ & $1.40 \times 10^{4}$ & $1.82 \times 10^{4}$ & 1 & $1.5 \times 10^{3}$ & $1.1 \times 10^{2}$ & 30 & 13 \\
$2.3 \times 10^{2}$ & \numrange{2.1e+02}{2.4e+02} & $\rm{M_R}$ & $1.26 \times 10^{4}$ & $1.82 \times 10^{4}$ & 1 & $1.7 \times 10^{3}$ & $1.3 \times 10^{2}$ & 34 & 15 \\
$2.6 \times 10^{2}$ & \numrange{2.4e+02}{2.8e+02} & $\rm{M_R}$ & $1.13 \times 10^{4}$ & $1.82 \times 10^{4}$ & 1 & $1.9 \times 10^{3}$ & $1.40 \times 10^{2}$ & 38 & 18 \\
$3 \times 10^{2}$ & \numrange{2.8e+02}{3.2e+02} & $\rm{M_R}$ & $1.01 \times 10^{4}$ & $1.82 \times 10^{4}$ & 1 & $2.1 \times 10^{3}$ & $1.9 \times 10^{2}$ & 42 & 21 \\
$3.5 \times 10^{2}$ & \numrange{3.2e+02}{3.8e+02} & $\rm{M_R}$ & $8.70 \times 10^{3}$ & $1.82 \times 10^{4}$ & 1 & $2.2 \times 10^{3}$ & $1.7 \times 10^{2}$ & 45 & 24 \\
$4 \times 10^{2}$ & \numrange{3.8e+02}{4.3e+02} & $\rm{M_R}$ & $7.68 \times 10^{3}$ & $1.82 \times 10^{4}$ & 1 & $2.5 \times 10^{3}$ & $2.1 \times 10^{2}$ & 49 & 28 \\
$4.7 \times 10^{2}$ & \numrange{4.3e+02}{5e+02} & $\rm{M_R}$ & $6.61 \times 10^{3}$ & $1.82 \times 10^{4}$ & 1 & $2.6 \times 10^{3}$ & $1.9 \times 10^{2}$ & 53 & 32 \\
$5.4 \times 10^{2}$ & \numrange{5e+02}{5.8e+02} & $\rm{M_R}$ & $5.88 \times 10^{3}$ & $1.82 \times 10^{4}$ & 1 & $2.9 \times 10^{3}$ & $2.4 \times 10^{2}$ & 58 & 38 \\
$6.2 \times 10^{2}$ & \numrange{5.8e+02}{6.7e+02} & $\rm{M_R}$ & $5.17 \times 10^{3}$ & $1.82 \times 10^{4}$ & 1 & $3.2 \times 10^{3}$ & $2.5 \times 10^{2}$ & 63 & 44 \\
$7.2 \times 10^{2}$ & \numrange{6.7e+02}{7.7e+02} & $\rm{M_R}$ & $4.34 \times 10^{3}$ & $1.82 \times 10^{4}$ & 1 & $3.3 \times 10^{3}$ & $2.8 \times 10^{2}$ & 67 & 50 \\
$8.3 \times 10^{2}$ & \numrange{7.7e+02}{8.9e+02} & $\rm{M_R}$ & $3.94 \times 10^{3}$ & $1.82 \times 10^{4}$ & 1 & $3.7 \times 10^{3}$ & $3.5 \times 10^{2}$ & 72 & 59 \\
$9.6 \times 10^{2}$ & \numrange{8.9e+02}{1e+03} & $\rm{M_R}$ & $3.27 \times 10^{3}$ & $1.82 \times 10^{4}$ & 1 & $3.8 \times 10^{3}$ & $4.8 \times 10^{2}$ & 79 & 67 \\
$1.1 \times 10^{3}$ & \numrange{1e+03}{1.2e+03} & $\rm{M_R}$ & $2.88 \times 10^{3}$ & $1.82 \times 10^{4}$ & 1 & $4.2 \times 10^{3}$ & $3.6 \times 10^{2}$ & 81 & 78 \\
$1.3 \times 10^{3}$ & \numrange{1.2e+03}{1.4e+03} & $\rm{M_R}$ & $2.37 \times 10^{3}$ & $1.82 \times 10^{4}$ & 1 & $4.3 \times 10^{3}$ & $4.4 \times 10^{2}$ & 87 & 88 \\
$1.5 \times 10^{3}$ & \numrange{1.4e+03}{1.6e+03} & $\rm{M_R}$ & $2.08 \times 10^{3}$ & $1.82 \times 10^{4}$ & 1 & $4.7 \times 10^{3}$ & $4.2 \times 10^{2}$ & 91 & $1.0 \times 10^{2}$ \\
$1.7 \times 10^{3}$ & \numrange{1.6e+03}{1.8e+03} & $\rm{M_R}$ & $1.66 \times 10^{3}$ & $1.82 \times 10^{4}$ & 1 & $4.6 \times 10^{3}$ & $5.6 \times 10^{2}$ & 97 & $1.1 \times 10^{2}$ \\
$2 \times 10^{3}$ & \numrange{1.8e+03}{2.1e+03} & $\rm{M_R}$ & $1.44 \times 10^{3}$ & $1.82 \times 10^{4}$ & 1 & $5.0 \times 10^{3}$ & $6.6 \times 10^{2}$ & 94 & $1.3 \times 10^{2}$ \\
$2.3 \times 10^{3}$ & \numrange{2.1e+03}{2.4e+03} & $\rm{M_R}$ & $1.09 \times 10^{3}$ & $1.82 \times 10^{4}$ & 1 & $4.7 \times 10^{3}$ & $6.8 \times 10^{2}$ & 98 & $1.4 \times 10^{2}$ \\
\end{longtable}
\end{ThreePartTable}

\section{Acknowledgements}

This paper is based (in part) on data obtained with the LOFAR telescope (LOFAR-ERIC). LOFAR (\citealt{vanHaarlem_2013}) is the Low Frequency Array designed and constructed by ASTRON. It has observing, data processing, and data storage facilities in several countries, that are owned by various parties (each with their own funding sources), and that are collectively operated by the LOFAR European Research Infrastructure Consortium (LOFAR-ERIC) under a joint scientific policy. The LOFAR-ERIC resources have benefited from the following recent major funding sources: CNRS-INSU, Observatoire de Paris and Université d'Orléans, France; Istituto Nazionale di Astrofisica (INAF), Italy; BMBF, MIWF-NRW, MPG, Germany; Science Foundation Ireland (SFI), Department of Business, Enterprise and Innovation (DBEI), Ireland; NWO, The Netherlands; The Science and Technology Facilities Council, UK; Ministry of Science and Higher Education, Poland.

This research made use of the Dutch national e-infrastructure with support of the SURF Cooperative (e-infra 180169) and NWO (grants 2019.056 \& 2023.036). The Jülich LOFAR Long Term Archive and the German LOFAR network are both coordinated and operated by the Jülich Supercomputing Centre (JSC), and computing resources on the supercomputer JUWELS at JSC were provided by the Gauss Centre for Supercomputing e.V. (grant CHTB00) through the John von Neumann Institute for Computing (NIC). 

This research made use of the University of Hertfordshire
high-performance computing facility and the LOFAR-UK computing
facility located at the University of Hertfordshire and supported by
STFC [ST/P000096/1], of the LOFAR-IT computing infrastructure
supported and operated by INAF, including the resources within the
PLEIADI special ``LOFAR'' project by USC-C of INAF, and by the Physics
Dept. of Turin University (under the agreement with Consorzio
Interuniversitario per la Fisica Spaziale) at the C3S Supercomputing
Centre, Italy. This work has been enabled by access to facilities and
the scientific and technical support provided by the UK SKA Regional
Centre (UKSRC). The UKSRC is a collaboration between the University of
Cambridge, University of Edinburgh, Durham University, University of
Hertfordshire, University of Manchester, University College London,
and the UKRI STFC Scientific Computing (STFC) at RAL. The UKSRC is
supported by funding from the UKRI STFC. Computing resources were made available via the federated Compute4PUNCH infrastructure established by the PUNCH4NFDI consortium. This publication also used resources from the LOFAR Data Valorization (LDV) projects [2020.031, 2022.033, 2024.047] of the research programme Computing Time on National Computer Facilities using SPIDER that is (co-)funded by the Dutch Research Council (NWO), hosted by SURF through the call for proposals of Computing Time on National Computer Facilities. 

M.J.H. thanks STFC for support under grants [ST/V000624/1] and [ST/Y001249/1]. A.D. acknowledges support by the German Federal Ministry of Research, Technology and Space (BMFTR) Verbundforschung under the grant 05A23STA. M.B. acknowledges support from the Deutsche Forschungsgemeinschaft under Germany's Excellence Strategy - EXC 2121 ``Quantum Universe'' - 390833306 and from the BMFTR ErUM-Pro grant 05A2023. J.R.C. acknowledges funding from the European Union via the European Research Council (ERC) grant Epaphus (project number 101166008). MH acknowledges funding from the ERC under the European Union's Horizon 2020 research and innovation programme (grant agreement No 772663). F.d.G., J.M.B., M.C., G.D.G., C.G., T.P. acknowledge support from the ERC Consolidator Grant ULU 101086378. L.K.M. is grateful for support from a UKRI FLF [MR/Y020405/1] and STFC support of LOFAR-UK [ST/V002406/1]. S.P.O. acknowledges support from the Comunidad de Madrid Atracci\'on de Talento program via grant 2022-T1/TIC-23797, and grant PID2023-146372OB-I00 funded by MICIU/AEI/10.13039/501100011033 and by ERDF, EU. D.J.S. acknowledges financial support from the BMFTR grant ErUM-Pro 05A23PB1. D.J.B.S. and J.C.S.P. acknowledge support from the UK's STFC under grants [ST/V000624/1]; H.K.V acknowledges funding from the European Research Council under the Horizon Europe programme of the European Union (grant no. 101042416 STORMCHASER) and from the Dutch research council (NWO) under the talent programme (Vidi grant VI.Vidi.203.093). G.J.W. gratefully acknowledges support of an Emeritus Fellowship from The Leverhulme Trust. L.A. acknowledges support from the UKRI STFC under the grant ST/X002543/1. B.B.-K. acknowledges support from INAF for the project \texttt{`Paving the way to radio cosmology in the SKA Observatory era: synergies between SKA pathfinders/precursors and the new generation of optical/near-infrared cosmological surveys'}, CUP C54I1900105 0001. M.Bi. and S.J.N. are supported by the Polish National Science Center through grant 2020/38/E/ST9/00395; M.Bi. also through 2020/39/B/ST9/03494. L.B. acknowledges support by the Studienstiftung des deutschen Volkes. M.Bri. acknowledges financial support from Next Generation EU funds within the National Recovery and Resilience Plan (PNRR), Mission 4 - Education and Research, Component 2 - From Research to Business (M4C2), Investment Line 3.1 - Strengthening and creation of Research Infrastructures, Project IR0000034 -- ``STILES - Strengthening the Italian Leadership in ELT and SKA'' and from INAF under the Mini Grant 2023 funding scheme (project `Low radio frequencies as a probe of AGN jet feedback at low and high redshift'). D.J.B., R.J.D., C.J.R. and M.S. acknowledge funding from the German Science Foundation DFG, via the Collaborative Reasearch Center SFB1491 ``Cosmic Interacting Matters - From Source to Signal''. A.B acknowledges financial support from the ERC CoG BELOVED n. 101169773. M.B. acknowledges support from the INAF minigrant ``A systematic search for ultra-bright high-z strongly lensed galaxies in Planck catalogues''. S.C. acknowledges support from the Italian Ministry of University and Research (\textsc{mur}), PRIN 2022 ``EXSKALIBUR -- Euclid-Cross-SKA: Likelihood Inference Building for Universe's Research'', Grant No.\ 20222BBYB9, CUP D53D2300252 0006, and from the European Union -- Next Generation EU. J.H.C. acknowledges the support of STFC via grants [ST/T000295/1] and [ST/X001164/1]. E.D.R. is supported by the Fondazione ICSC, Spoke 3 Astrophysics and Cosmos Observations. National Recovery and Resilience Plan (Piano Nazionale di Ripresa e Resilienza, PNRR) Project ID CN\_00000013 ``Italian Research Center for High-Performance Computing, Big Data and Quantum Computing'' funded by MUR Missione 4 Componente 2 Investimento 1.4: Potenziamento strutture di ricerca e creazione di ``campioni nazionali di R\&S (M4C2-19)'' -- Next Generation EU (NGEU). J.M.G.H.J.d.J. acknowledges support from project CORTEX (NWA.1160.18.316) of research programme NWA-ORC, which is (partly) financed by the Dutch Research Council (NWO), and support from the OSCARS project, which has received funding from the European Commission's Horizon Europe Research and Innovation programme under grant agreement No. 101129751. R.J.D. acknowledges support from BMFTR under grant 05A23PC2. K.J.D acknowledges support from the STFC through an Ernest Rutherford Fellowship (grant number ST/W003120/1). M.J.J. acknowledges the support of a UKRI Frontiers Research Grant [EP/X026639/1], which was selected by the European Research Council the STFC consolidated grants [ST/S000488/1] and [ST/W000903/1]. M.J.J and C.L.H. acknowledge support from the Oxford Hintze Centre for Astrophysical Surveys which is funded through generous support from the Hintze Family Charitable Foundation. M.K.  MK acknowledges acknowledge support from the Deutsche Forschungsgemeinschaft (DFG, grant 443220636 [FOR5195: Relativistic Jets in Active Galaxies])and from the BMBF ErUM-Pro grant 05A2023. R.K. acknowledges support from the Leverhulme Trust Early Career Fellowship. M.K.B. acknowledges support from the Polish National Science Centre under grant no. 2017/26/E/ST9/00216. M.M. and I.P. acknowledge support from INAF under the Large Grant 2022 funding scheme (project ``MeerKAT and LOFAR Team up: a Unique Radio Window on Galaxy/AGN co-Evolution''). K.M. acknowledges support from the Polish National Science Centre, NCN 2024/53/B/ST9/00230. J.P.M acknowledges support from the National Research Foundation of South Africa (Grant Number: 128943). B.M. acknowledges support from UKRI STFC for an Ernest Rutherford Fellowship (grant number ST/Z510257/1). J.M. acknowledges financial support from the Severo Ochoa grant CEX2021-001131-S and grant PID2023-147883NB-C21, funded by MCIU/AEI/ 10.13039/501100011033. D.G.N. acknowledges funding from Conicyt through Fondecyt Postdoctorado (project code 3220195), Comit\'e Mixto ESO-Chile, and N\'ucleo Milenio TITANs (projects NCN2022002, NCN23\_002). M.P.A. acknowledges support from the Bundesministerium f\"ur Bildung und Forschung (BMFTR) ErUM-IFT 05D23PB1. I.D.R. acknowledges support from the Banting Fellowship Program. F.S. appreciates the support of STFC [ST/Y004159/1]. R.T. is grateful for support from the UKRI Future Leaders Fellowship (grant MR/T042842/1). M.V. acknowledges financial support from the Inter-University Institute for Data Intensive Astronomy (IDIA), a partnership of the University of Cape Town, the University of Pretoria and the University of the Western Cape, and from the South African Department of Science and Innovation's National Research Foundation under the ISARP RADIOMAP Joint Research Scheme (DSI-NRF Grant Number 150551) and the CPRR HIPPO Project (DSI-NRF Grant Number SRUG22031677).

\twocolumn

\end{appendix}

\end{document}